\documentclass[prx,10pt,superscriptaddress,twocolumn,aps,showpacs,floatfix,longbibliography]{revtex4-2}

\usepackage{xr}
\usepackage{amsmath}
\usepackage{amsfonts, amssymb,amsxtra, physics}
\usepackage[]{graphicx}
\usepackage{grffile}
\usepackage{framed}
\usepackage{bbm}
\usepackage{braket}
\usepackage[dvipsnames]{xcolor}
\usepackage{chngcntr}
\usepackage{comment}
\usepackage[normalem]{ulem}

\counterwithout{equation}{section} 

\usepackage{bm}
\usepackage{epsfig}
\usepackage{blindtext}
\usepackage{floatrow}
\usepackage{color}
\usepackage{diagbox}
\usepackage{outlines}
\usepackage{cancel}
\usepackage{dsfont}
\usepackage{soul}
\usepackage{float}

\usepackage[colorlinks=true]{hyperref}

\begin{document}
\title{Concentration-Free Quantum Kernel Learning in the Rydberg Blockade}
\author{Ayana Sarkar}
    \altaffiliation{These authors contributed equally to this work.\\ \textcolor{magenta}{ayana.sarkar@usherbrooke.ca}\\ \textcolor{magenta}{martin.schnee@usherbrooke.ca}}
    \affiliation{Institut quantique, Sherbrooke, Québec, J1K 2R1, Canada}
    \affiliation{Département de physique, Université de Sherbrooke, Sherbrooke, Québec, J1K 2R1, Canada}
\author{Martin Schnee}
    \altaffiliation{These authors contributed equally to this work.\\ \textcolor{magenta}{ayana.sarkar@usherbrooke.ca}\\ \textcolor{magenta}{martin.schnee@usherbrooke.ca}}
    \affiliation{Institut quantique, Sherbrooke, Québec, J1K 2R1, Canada}
    \affiliation{Département de physique, Université de Sherbrooke, Sherbrooke, Québec, J1K 2R1, Canada}  
\author{Sangeeth Das Kallullathil}
    \affiliation{Institut quantique, Sherbrooke, Québec, J1K 2R1, Canada}
    \affiliation{Département de physique, Université de Sherbrooke, Sherbrooke, Québec, J1K 2R1, Canada}  
\author{Roya Radgohar}
    \affiliation{Institut quantique, Sherbrooke, Québec, J1K 2R1, Canada}
    \affiliation{Département de physique, Université de Sherbrooke, Sherbrooke, Québec, J1K 2R1, Canada}
\author{Mojde Fadaie}
    \affiliation{Institut quantique, Sherbrooke, Québec, J1K 2R1, Canada}
    \affiliation{Département de physique, Université de Sherbrooke, Sherbrooke, Québec, J1K 2R1, Canada}
\author{Victor Drouin-Touchette}
    \affiliation{Institut quantique, Sherbrooke, Québec, J1K 2R1, Canada}
    \affiliation{Département de physique, Université de Sherbrooke, Sherbrooke, Québec, J1K 2R1, Canada}
    \affiliation{Département de génie électrique et de génie informatique, Université de Sherbrooke, Sherbrooke, Québec, J1K 2R1, Canada}
\author{Stefanos Kourtis}
    \affiliation{Institut quantique, Sherbrooke, Québec, J1K 2R1, Canada}
    \affiliation{Département de physique, Université de Sherbrooke, Sherbrooke, Québec, J1K 2R1, Canada}
    \affiliation{Département d'informatique, Université de Sherbrooke, Sherbrooke, Québec, J1K 2R1, Canada}

\date{\today}

\begin{abstract}
Quantum kernel methods (QKMs) offer an appealing framework for machine learning on near-term quantum computers.
However, QKMs generically suffer from exponential concentration, requiring an exponential number of measurements to resolve kernel values, with the exception of trivial (i.e., classically simulable) kernels.
Here we propose a QKM that is free of exponential concentration, yet remains hard to simulate classically. Our QKM utilizes the weak ergodicity-breaking many-body dynamics in the Rydberg blockade of coherently driven neutral atom arrays. 
We demonstrate the fundamental properties of our QKM by analytically solving an approximate toy model of its underpinning quantum dynamics, as well as by extensive numerical simulations on randomly generated datasets. 
We further show that the proposed kernel exhibits effective learning on real data. 
The proposed QKM can be implemented in current neutral atom quantum computers. Along the way, we uncover novel physical insights into the thermalization of weak ergodicity-breaking systems through the \emph{non-stabilizerness} of the underlying Rydberg-blockaded dynamics, which directly governs the classical simulability of the proposed kernel.
\end{abstract}

\maketitle
\section{Introduction}
Designing quantum algorithms that (1) address practically relevant tasks, (2) remain hard to simulate classically, and (3) are compatible with the limited capabilities of noisy quantum hardware represents one of the central challenges in quantum science today, and a prerequisite for achieving a quantum advantage, namely, a computational speedup over the best classical method for the solution of a given problem. Machine learning (ML), in particular, is touted as a prime application where near-term quantum computing can fulfill these three criteria~\cite{Perdomo-Ortiz_2018}, targeting problems of academic or commercial relevance beyond those artificially tailored to specific quantum hardware. In the last years, several results have established a separation between the capabilities of quantum machine learning (QML) models, taking the form of parameterized quantum unitaries, and their classical counterparts for some specific tasks~\cite{kubler2021inductivebiasquantumkernels, QKL-ECfree-Huang2021,Huang2022,Liu2021}. 

We focus on kernel methods, a well established class of supervised learning techniques that project data into high-dimensional feature spaces~\cite{Kernel-ScholkopfSmola, Kernel-Shawe-Taylor_Cristianini_2004}, and extend naturally to the quantum domain. In quantum kernel methods (QKMs), data points are embedded into the Hilbert space of an ensemble of qubits~\cite{Schuld2019,Schuld2021,kubler2021inductivebiasquantumkernels}, after which their similarity can be evaluated by measuring fidelity between quantum states. The kernel matrix storing this similarity information for all pairs of points in the training dataset can then be used for classification or regression tasks in a classical convex optimization scheme with convergence guarantees. The promise of near-term applicability has sparked extensive studies of the promises and challenges of QKMs~\cite{Schuld2021, schuld2021supervisedquantummachinelearning, thanasilp2024exponential, kubler2021inductivebiasquantumkernels,kairon2025equivalenceexponentialconcentrationquantum, Mcclean2018BP, cerezoVQA2021, cerezo2023, diaz2023, Schnabel_2025, QEK-Hen2021, QEK-Henriet2023}. 

However, QKMs generically suffer from exponential concentration (EC): off-diagonal kernel elements vanish exponentially as the number of qubits $N$ increases, thus requiring an exponential number of measurements to precisely evaluate the kernel values on a quantum computer~\cite{thanasilp2024exponential,kubler2021inductivebiasquantumkernels,kairon2025equivalenceexponentialconcentrationquantum,Mcclean2018BP,cerezoVQA2021,cerezo2023,diaz2023}. This problem is akin to the issue of barren plateaus (BPs) in variational quantum algorithms~\cite{BP-McClean2018,BP-QNN2018-2,BP-GWOB2019,BP-ACCCC2021,BP-CSVCC2021,BP-QWZGG2023,BP-larocca2024reviewvariational, Arrasmith_2022_QBP}. While numerous strategies have been developed to mitigate BPs~\cite{BP-free-10485449,BP-free-haug2021optimaltrainingvariationalquantum,BP-free-PhysRevA.106.L060401,BP-free-PRXQuantum.3.020365,BP-Free-PhysRevResearch.6.013076,BP-free-Sannia2024,BP-free-Grimsley2023,BP-free-shi2024avoidingbarrenplateausgaussian}, efforts to address EC of quantum kernels remain limited~\cite{QKL-ECFree-PhysRevA.106.042407,QKL-ECfree-suzuki2022quantumfisherkernelmitigating,QKL-EC-free-Glick2024,QKL-ECfree-Huang2021,kairon2025equivalenceexponentialconcentrationquantum, saem2025pitfallstacklingexponentialconcentration, nakaji2021expressibility}; these have mainly focused on mitigating EC effects in existing QKMs~\cite{QKL-ECfree-Huang2021,QKL-ECFree-PhysRevA.106.042407, QKL-EC-free-Glick2024, QKL-ECfree-suzuki2022quantumfisherkernelmitigating, saem2025pitfallstacklingexponentialconcentration}. Moreover, many proposals are hard to implement on currently available quantum hardware hindering the direct assessment of the concentration issue~\cite{Zimboras2025}.

At the same time, advances in the experimental control of individual quantum systems have enabled analog quantum simulators to operate in regimes where quantum advantage may be within reach, even in the presence of noise~\cite{Daley2022,leclerc2026}. In particular, non-equilibrium dynamics occurring in programmable atomic arrays operated in the strongly-interacting Rydberg-blockaded regime are capable of reaching high-entanglement entropy regimes \cite{Shaw2024} and producing complex constrained many-body dynamics in one and two dimensions \cite{bernien2017probing,bluvstein2021controlling}. 

\begin{figure}[!tbp]
\includegraphics[width=\textwidth]{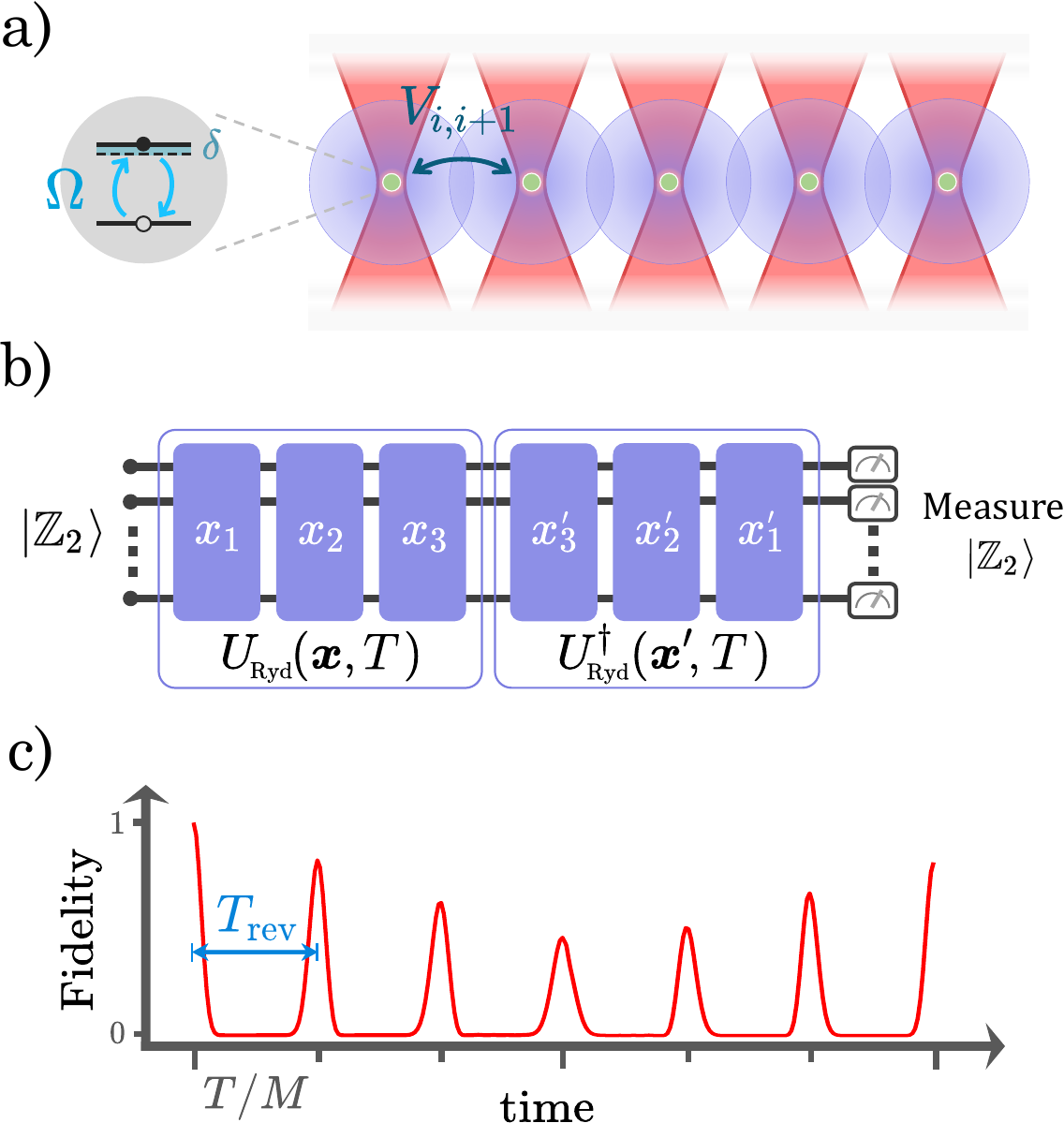} 
\caption{Schematic illustration of \textsc{RydKernel}. (a) Neighboring laser-trapped Rydberg atoms can be subjected to constrained Rydberg-blockaded interactions. (b) Instance of the \textsc{RydKernel} circuit (echo protocol) for a pair of data points $(\boldsymbol{x},\boldsymbol{x'})$ with $M=3$ features and encoding time $T=3T_{\text{rev}}$. (c) \textsc{RydKernel} viewed as initial $\ket{\mathbb{Z}_2}$ state fidelity in time shows the long-lived periodic many-body revivals relative to which encoding time is expressed in our setup.}
\label{fig:illustration}
\end{figure}

In this work, we propose \textsc{RydKernel}, a QKM that is inherently free from EC. \textsc{RydKernel} leverages the weak ergodicity-breaking many-body dynamics native to coherently driven neutral-atom arrays. In this construction, data is encoded in the detuning perturbation of the driving laser and the constrained dynamics is used to promote large state overlaps and avoid EC (see schematic illustration in Fig.\ref{fig:illustration}). In addition, \textsc{RydKernel} is hard to simulate classically and implementable on current analog neutral atom quantum computers (NAQC). As such, we show that \textsc{RydKernel} fulfills the prerequisite criteria (1-3) for a potential quantum advantage stated in the first paragraph. 

The remainder of this paper is organized as follows. Sec.~\ref{expoconcdef} covers quantum kernel methods and exponential concentration. Sec.~\ref{RydKernel} introduces \textsc{RydKernel} and the Rydberg-blockaded dynamics underpinning it. In Sec.~\ref{toykernelfull}, we discuss a toy model of the underlying dynamics, treating the evolution as a large precessing spin and characterizing the concentration behavior of the kernel analytically. This toy model allows us to establish that for small encoding perturbations the kernel mean and variance scale polynomially in the system size, thus rigorously precluding EC. These analytical predictions are confirmed by extensive numerical simulations of \textsc{RydKernel} on randomly generated datasets. Furthermore, we show numerically that the anti-concentration properties of our kernel persist for relatively stronger encoding perturbations and identify a critical value beyond which EC is recovered, delineating the boundaries of the EC-free phase of \textsc{RydKernel}. This is the main result of this paper (Sec~\ref{resultsmain}). 

Additionally, we provide evidence for the classical hardness of simulating \textsc{RydKernel} through complementary arguments. First in Sec.~\ref{entngrowth}, we show that the dynamics quickly generates volume-law entanglement, a bottleneck for tensor network methods. Second in Sec.~\ref{magicgrowth}, we lower-bound the growth of (non-local) magic in the system via its relation to the anti-flatness of the entanglement spectrum, and find that it exhibits system-size scaling at intermediate times, a hard limit for any stabilizer-based simulation of the kernel. Along the way, we uncover a large “magic barrier”, offering novel insights on thermalization in weak ergodicity-breaking dynamics. 
Finally, in Sec.~\ref{MLperf} we demonstrate the generalization capacity of \textsc{RydKernel} on the IRIS classification benchmark and discuss its implementation on current analog neutral atom quantum computers or its simulation on fault-tolerant digital quantum architectures in Sec.~\ref{expimplmntation}. While we do not show an advantage over classical algorithms for the tested learning task, we emphasize the importance of introducing to the community a framework that fulfills the quantum advantage prerequisites.

Overall, \textsc{RydKernel} offers a practical framework for meaningful near-term quantum ML on existing hardware, and more broadly, highlights the potential of using constrained many-body dynamics as a resource for near-term quantum algorithms, opening doors to quantum advantage in QML.

\section{Exponential Concentration in Quantum Kernel Methods}
\label{expoconcdef}
\subsection{Quantum Kernel Methods}

Let us begin by defining quantum kernel methods. First, consider a classical training data set $\mathcal{S}$ formed of $N_s$ data vectors $\boldsymbol{x} \in \mathcal{X}$, each associated to its class label $y \in \mathcal{Y}$. Without loss of generality, we can choose $\mathcal{X} = [0,1]^M$, i.e., data vectors composed of $M$ normalized real features, and $\mathcal{Y} = \{0,1\}$, corresponding to a binary classification task. 

In QKMs, data vectors $\boldsymbol{x}$ are embedded in a parameterized unitary $U(\boldsymbol{x})$ used to obtain a quantum state $|\psi(\boldsymbol{x})\rangle = U(\boldsymbol{x})|\psi_0\rangle$, with $|\psi_0\rangle$ a fixed initial product state on $N$ qubits (typically, $|\psi_0\rangle = |0\rangle^{\otimes N} \equiv |\mathbf{0}\rangle$). This data-encoding unitary induces a data-dependent ensemble of quantum states whose geometry governs both the expressive power and the practical trainability of the resulting kernel.
The fidelity quantum kernel is defined as
\begin{equation}
    \kappa (\boldsymbol{x}, \boldsymbol{x'}) = |\langle \psi(\boldsymbol{x}) | \psi(\boldsymbol{x'})\rangle|^2 
    \label{eq:kernel}
\end{equation}
and measures the similarity between two data points $\boldsymbol{x}$ and $\boldsymbol{x'}$ via the inner product of their corresponding quantum states. Evaluating $\kappa(\boldsymbol{x},\boldsymbol{x}')$ amounts to applying the unitary sequence $U^\dagger(\boldsymbol{x})\,U(\boldsymbol{x}')$ on the initial state $|\psi_0\rangle$ and estimating the probability of measuring $|\psi_0\rangle$ — a quantity directly accessible as a Loschmidt echo~\cite{GPSZLoschEcho2006}. \\

The kernel values populate an $N_s \times N_s$ Gram matrix with entries $\kappa_{ij} = \kappa(\boldsymbol{x}_i, \boldsymbol{x}_j)$ for all pairs of data points. This kernel matrix is used in combination with the class labels $\mathcal{Y}$ to build a loss function. The convexity of the loss landscape guarantees classical convergence towards an optimal classification model~\cite{Schuld2021,schuld2021supervisedquantummachinelearning,thanasilp2024exponential} for a given kernel. Note that this is in stark contrast to variational quantum approaches where non-convexity can restrict trainability. 

\subsection{Exponential Concentration}
\label{definitionEC}
A central obstacle to the scalable deployment of QKMs on quantum computers is the problem of exponential concentration of kernel values. 

Quantum devices being inherently probabilistic, kernel values $\kappa(\boldsymbol{x},\boldsymbol{x}')$ can only be statistically estimated through a finite number of measurements. As the number of qubits $N$ grows, kernel values may become exponentially small and almost constant, making them indistinguishable from one another within any finite measurement (shot) budget. Resolving such kernels beyond statistical noise would require an exponentially large number of shots, rendering them practically useless for learning.

More precisely, the quantum kernel $\kappa$ is said to be \emph{deterministically exponentially concentrated} towards a value
$\mu = \mathbb{E}_{\boldsymbol{x},\boldsymbol{x}'}
[\kappa(\boldsymbol{x},\boldsymbol{x}')]$
if
\begin{equation}
    \bigl|\kappa(\boldsymbol{x},\boldsymbol{x}') - \mu\bigr|
    \leq \beta, \quad \beta \in \mathcal{O}(b^{-N}),
    \label{eq:EC_det}
\end{equation}
for some $b > 1$ and all pairs $(\boldsymbol{x}, \boldsymbol{x}')$.
More generally, $\kappa$ is
\emph{probabilistically exponentially concentrated} if
\begin{equation}
    \mathrm{Pr}_{\boldsymbol{x},\boldsymbol{x}'}
    \!\left[\,\bigl|\kappa(\boldsymbol{x},\boldsymbol{x}')
    - \mu\bigr| \geq \delta\right]
    \leq \frac{\beta}{\delta^2},
    \quad \beta \in \mathcal{O}(b^{-N}),
    \label{eq:EC_prob}
\end{equation}
for some $b>1$, so that the probability of observing any kernel entry
that deviates from $\mu$ by a fixed amount $\delta > 0$ becomes
exponentially small in $N$. Probabilistic EC can be equivalently diagnosed by studying the variance of the kernel over the data distribution \cite{thanasilp2024exponential}
\begin{equation}
\mathrm{Var}_{\boldsymbol{x},\boldsymbol{x}'}\!\left[\kappa(\boldsymbol{x},\boldsymbol{x}')\right]
    \in \mathcal{O}(b^{-N}), \quad b > 1.
    \label{eq:EC_var}
\end{equation}
Note that if $\mu$ additionally vanishes exponentially with $N$, i.e.\
$\mu \in \mathcal{O}(b'^{-N})$ for some $b' > 1$, the kernel concentrates
towards an exponentially small value.
In either case, in presence of EC, all kernel entries collapse toward the
same mean value $\mu$, so that $\kappa(\boldsymbol{x}_i,\boldsymbol{x}_j)
\approx \mu$ for all $i,j$.
The kernel (Gram) matrix then carries no information about pairwise similarities
between data points, and any model trained on it produces predictions
that are independent of the input data~\cite{thanasilp2024exponential}.
An exponentially large number of measurements — growing as
$\mathcal{O}(b^N)$ — would be required to resolve kernel entries from
their mean, which is prohibitive on any near-term quantum device.
EC therefore simultaneously prevents efficient kernel estimation
\emph{and} destroys generalisation, rendering the QKM practically
useless beyond small system sizes.

\subsection{Connection to Barren Plateaus}
\label{connectionBP}
EC is closely related to the problem of \textit{barren plateaus} (BPs) in variational quantum algorithms 
(VQAs)~\cite{BP-McClean2018, cerezoVQA2021, BP-larocca2024reviewvariational}. It is worth noting the distinction between EC in QKMs and BPs
in VQAs, as the two phenomena are related but not identical. 

In the VQA setting, a cost function is said to exhibit a BP if the variance of its gradient with respect to the
variational parameters vanishes exponentially with the number of qubits 
\cite{BP-McClean2018,cerezoVQA2021,BP-larocca2024reviewvariational}.
Note that, where EC concerns the concentration of kernel values over the \textit{data distribution}, BPs concern the concentration of cost
function gradients over the \textit{parameter space} — two distinct
variances that nevertheless share the same exponential suppression
mechanism. 

Recently, Ref. \cite{kairon2025equivalenceexponentialconcentrationquantum} formalized this connection into a rigorous equivalence theorem, showing that if a parameterized ansatz induces a BP landscape in the VQA setting,  then the quantum kernel constructed from the same circuit concentrates exponentially, and vice versa. Consequently, the variance bounds derived  for cost functions in the BP literature transfer directly to bounds on  kernel matrix entries.

Despite the severity of the problem of EC, dedicated strategies for avoiding it remain far less developed than the corresponding BP mitigation literature~\cite{BP-McClean2018,BP-QNN2018-2,BP-GWOB2019,BP-ACCCC2021,BP-CSVCC2021,BP-QWZGG2023,BP-larocca2024reviewvariational}. Approaches proposed so far include projected quantum kernels~\cite{QKL-ECfree-Huang2021}, bandwidth 
rescaling~\cite{QKL-ECFree-PhysRevA.106.042407}, quantum Fisher 
kernels~\cite{QKL-ECfree-suzuki2022quantumfisherkernelmitigating}, and 
symmetry-constrained covariant 
kernels~\cite{QKL-EC-free-Glick2024}. A common underlying principle in designing EC-free models is the introduction of structural inductive bias to restrict the effective Hilbert space dimension explored by the model \cite{kubler2021inductivebiasquantumkernels, saem2025pitfallstacklingexponentialconcentration}. 

\subsection{Sources of Exponential Concentration}
\label{sourceEC}
The origin of EC for fidelity quantum kernels can be traced to three mechanisms that can each individually lead to concentration in kernels \cite{thanasilp2024exponential}.

\textit{Firstly}, the expressivity of the data-encoding ensemble
$\{U(\boldsymbol{x})\}_{\boldsymbol{x} \in \mathcal{X}}$ — defined as
how uniformly it covers the unitary group $\mathrm{U}(2^N)$ as
$\boldsymbol{x}$ varies — plays a central role in driving EC.
Since the kernel in Eq.~\eqref{eq:kernel} is an inner product between two
states in a $2^N$-dimensional Hilbert space, a highly expressive
encoding renders $|\psi(\boldsymbol{x})\rangle$ and
$|\psi(\boldsymbol{x}')\rangle$ effectively random vectors in this exponentially large space.
Two such vectors are nearly orthogonal with high probability, causing kernel values to concentrate around zero~\cite{thanasilp2024exponential}.
This exposes an inherent tension in QKM design: the expressivity that enriches the feature space simultaneously destroys the data dependence
of the kernel. Avoiding EC therefore demands encodings that are structured enough to
preserve non-trivial, data-dependent overlaps at scale. 

\textit{Secondly}, the global nature of the measurement required to evaluate the fidelity kernel (Eq.~\eqref{eq:kernel}) constitutes an independent source of EC. Estimating the return probability to the initial state involves a
measurement that acts non-trivially on all $N$ qubits simultaneously, attempting to extract a single scalar from a state supported on a $2^N$-dimensional Hilbert space. 

\textit{Thirdly}, hardware noise presents a further, and in many respects the most practically significant, source of EC. Unlike the preceding sources, hardware noise is not an artefact of algorithmic design but is intrinsic to the quantum device or the associated measurement process, and its precise character depends strongly on the platform.

Therefore, designing a quantum kernel that avoids EC while retaining a genuine quantum advantage constitutes an open challenge and is the central motivation of the present work. In the sections that follow, we introduce a quantum kernel based on the Rydberg blockade and provide both analytical and numerical evidence that it is free from EC despite relying on the measurement of a global observable and scrambling in an exponentially large Hilbert space.

\section{Quantum kernel based on Rydberg blockade}
\label{RydKernel}

In this section we introduce our quantum kernel, which we call \textsc{RydKernel}. It is based on the Rydberg-blockaded quantum dynamics arising in analog Rydberg-atom quantum computers.

A neutral-atom quantum computer (NAQC) consists of alkali atoms trapped in real-space using optical tweezers and cooled to their electronic ground state.
For example, we can consider Rb atoms, whose ground state is $|g\rangle = |5S_{1/2}\rangle$.
A laser with frequency $\omega = \omega_{gr} + \delta$ then drives atoms to the Rydberg state $|r\rangle = |60S_{1/2}\rangle$, where $\omega_{gr}$ is the resonant frequency between $|g\rangle$ and $|r\rangle$, and $\delta$ is the detuning. 
Strong dipole-dipole interactions between nearby atoms in the Rydberg state gives rise to the Rydberg blockade mechanism, which produces intricate entanglement~\cite{Shaw2024} and powers analog quantum computation and simulation~\cite{wintersperger2023neutral,Browaeys2020}.

\subsection{\textsc{RydKernel}}

We consider a register of $N$ trapped atoms arranged in a chain with open boundary conditions, described by evolution under the Rydberg Hamiltonian ($\hbar=1$)
\begin{equation}
    H_{\text{Ryd}}(\lambda x) = \frac{\Omega}{2} \sum_{i=1}^{N} \bigg( \sigma_i^x - \lambda x \hat{n}_i \bigg) + \sum_{i<j} V_{ij} \hat{n}_i \hat{n}_j \; ,
    \label{eq:rydberg}
\end{equation}
where $\sigma_i^x = |r\rangle_i\langle g|_i + |g\rangle_i\langle r|_i$, $\hat{n}_i = |r\rangle_i\langle r|_i$, $\Omega$ is the local Rabi oscillation frequency. 
Atoms at sites $i$ and $j$ interact via van der Waals interactions $V_{ij} \approx C_6/r_{ij}^{6}$, with $C_6 >0$ and $r_{ij}$ the real-space distance between atoms, effective only if both atoms are in the Rydberg state.
The laser detuning $\delta$ is parametrized as $\lambda x = 2\delta/\Omega $, where $x \in [0,1]$ are rescaled data vector components.
This parametrization emphasizes that we use the renormalized detuning $\lambda x$ in our kernel for encoding data.

In \textsc{RydKernel}, we harness the constrained dynamics arising from the strongly-interacting Rydberg blockade regime, $V_{i, i+1} \gg \Omega \gg V_{i, i+2}$, and the Néel state $|\mathbb{Z}_2\rangle = |rgrg\cdots\rangle$ to build a quantum kernel that is free from EC. For $\boldsymbol{x}, \boldsymbol{x'} \in \mathcal{X}$, the kernel is defined as
\begin{equation}
    \kappa_{\text{Ryd}} (\boldsymbol{x}, \boldsymbol{x'}) = \big|\langle \mathds{Z}_{2} | U_{\text{Ryd}}^{\dagger}(\lambda\boldsymbol{x}, T) U_{\text{Ryd}}(\lambda \boldsymbol{x'}, T) |\mathds{Z}_{2} \rangle \big|^2 \,,
    \label{eq:rydbergkernel}
\end{equation}
where the parameterized unitary evolution is
\begin{equation}
    U_{\text{Ryd}}(\lambda\boldsymbol{x}, T) = \prod_{m=1}^{M} \exp\left[-i \frac{T}{M} H_{\text{Ryd}}(\lambda x_m)  \right] .
    \label{eq:encodingunitary}
\end{equation}
The components $\{x_m\}_{m=1,...,M}$ of a given point $\boldsymbol{x}$ (resp. $\{x'_{m'}\}_{m'=1,...,M}$ for $\boldsymbol{x'}$) are encoded in layers of successive evolution through the detuning strength $x_m$ of the Rydberg Hamiltonian of Eq.~\eqref{eq:rydberg}. The total encoding time is $T$ and the overall magnitude of the detuning is $\lambda$.

\subsection{Rydberg-blockaded dynamics}

We focus on the strongly-interacting Rydberg blockade regime $V_{i, i+1} \gg \Omega \gg V_{i, i+2}$, where pairs of atoms within a distance smaller than the blockade radius $r_B = (C_6/\Omega)^{1/6}$ are energetically prohibited to simultaneously transition from ground state to Rydberg state.

In this regime, it is known that the low-energy sector of Eq.~\eqref{eq:rydberg} can be described by the effective PXP Hamiltonian, 
\begin{align}
    H_{\text{PXP}}(\lambda x) = \frac{\Omega}{2}\sum_{i=1}^N P_{i-1} \left( \sigma_i^x - \lambda x \hat{n}_i \right) P_{i+1} \,,
    \label{eq:PXP}
\end{align} 
with local ground-state projector $P_i = \mathds{1}-\hat{n}_i = (\mathds{1}-\sigma_i^z)/2$, where the Rydberg blockade constraints are explicit.

These constraints restrict the nonintegrable quantum dynamics within a Hilbert subspace of exponential size ($\text{dim}\sim \phi^N$, with $\phi$ the golden ratio). 
It is also well known that, if the system is initialized in the Néel state $|\mathbb{Z}_2\rangle = |rgrg\cdots\rangle$~\footnote{This also occurs for specific initial product states \cite{turner2018quantum, turner2018weak, Serbyn2021}.}, the many-body dynamics at zero detuning ($\lambda = 0$) gives rise to revivals of period $T_{\text{rev}}\simeq 1.38 \times 2\pi/\Omega$ independent of system size. 
These have been observed experimentally~\cite{bernien2017probing} and are long-lived ~\cite{turner2018quantum, turner2018weak, Serbyn2021}. 
Systems obeying the effective PXP Hamiltonian host weak-ergodicity breaking dynamics, i.e., revivals, in one and two dimensions~\cite{bernien2017probing,bluvstein2021controlling}. This phenomenon stems from the existence of $N+1$ equidistant low-entangled energy eigenstates, dubbed \textit{quantum many-body scars} (QMBS), in the otherwise thermal spectrum of the Hamiltonian of Eq. ~\eqref{eq:PXP}. Although the $|\mathbb{Z}_2\rangle$ initial state has prominent overlap with these QMBS states, it also has support on an exponential number of energy eigenstates.

Finally, the magnitude of the detuning used to encode data can be chosen to be a small perturbation to the resonant drive, $\lambda \ll 1$, ensuring the revival dynamics is preserved \cite{turner2018quantum}. If strong enough, this perturbation changes the frequency of oscillations and induces dephasing by removing the periodic energy spacing between the QMBS eigenstates. However, QMBS themselves are preserved in the energy spectrum at least up to $\lambda=1$, as well as their high overlap with initial $|\mathbb{Z}_2\rangle$ state.

\section{Toy kernel}
\label{toykernelfull}
To build intuition on the behavior of the kernel in Eq. \eqref{eq:rydbergkernel}, we draw insights from a toy model of the weak-ergodicity breaking dynamics of the initial $|\mathbb{Z}_2\rangle$ state. 
At early times, this dynamics is quasi-integrable and obeys an approximate $\text{SU}(2)$ algebra ~\cite{choi2019emergent}. 
Thus, it can be approximately described by a large spin $\vec{S}$.

We use it to build an integrable toy kernel for which a perturbative closed-form expression, Eq. \eqref{eq:toykernelscaling}, can be analytically derived in the case of single feature encoding. This leads to analytical expressions of mean and variance for random datasets, Eqs. \eqref{eq:toykernelmean} and \eqref{eq:toykernelvar}, predicting an absence of exponential concentration. 

\subsection{Toy kernel from an approximate SU(2) toy model of the dynamics}
\label{toymodephysics}
It was shown in \cite{Turner_2018, choi2019emergent} that the oscillatory dynamics unfolding from the initial $|\mathbb{Z}_2\rangle$ state can be described at early times by the forward scattering approximation (FSA). In this view, the PXP Hamiltonian at zero detuning ($\lambda=0$) is represented as
\begin{equation}
    H_{\text{PXP}} = H^+ + H^- \; ,
\end{equation}
with $H^{\pm} = \sum_{j \text{ even}} P_{j-1} \sigma_{j}^{\pm} P_{j+1}  + \sum_{j \text{ odd}} P_{j-1} \sigma_{j}^{\mp} P_{j+1}$, where $\sigma_{j}^{\pm}$ are raising and lowering operator of the staggered magnetization $S^{\pi} = \frac{1}{2} \sum_i (-1)^i \sigma_i^z$, such that $[S^{\pi}, H^{\pm}]=\pm H^{\pm}$ and $H^+ = (H^-)^{\dagger}$. 

Moreover, the $\ket{\mathbb{Z}_2}$ state is the maximum-value eigenstate of $S^{\pi}$. This suggests that the dynamics can be viewed as a precessing large spin $\vec{S}$. However, the algebra generated by $H^{\pm}$ and $S^{\pi}$ is only an approximate $\text{SU}(2)$ algebra since $[H^+, H^-]=\frac{1}{2} S^{\pi} + \frac{1}{4} O_{ZZZ}$, where $O_{ZZZ} = \sum_i (-1)^i \sigma_{i-1}^z \sigma_i^z \sigma_{i+1}^z$.\\

This motivates us to consider the following $N$-qubit collective-spin Hamiltonian as an integrable toy model for the early time dynamics, 
\begin{equation}
    H_{\text{toy}} (\lambda x) =  \frac{\Omega_{\text{toy}}}{2} \left(S^{x} + \lambda x V \right) \,,
    \label{eq:toymodel}
\end{equation}
where $S^{x} = \sum_{i = 1}^{N} \sigma^{x}_{i}$. When initialized in the maximum-eigenvalue eigenstate $|\psi_0\rangle = |\boldsymbol{0} \rangle$ of $S^z$ (with the correspondence $\ket{r} \leftrightarrow \ket{0}$, $\ket{g} \leftrightarrow \ket{1}$ for the single qubit basis), the large spin $\vec{S}$ precesses at frequency $\Omega_{\text{toy}} = \Omega/1.38$. The perturbation $V = \sum_{i}^{N}(-1)^{i} \sigma_{i}^{z}$ here plays the same role regarding $|\boldsymbol{0}\rangle$ as the detuning perturbation does for $\ket{\mathbb{Z}_2}$ in the Rydberg dynamics. We construct a toy kernel $\kappa_{\text{toy}}$ by replacing $H_{\text{Ryd}}(\lambda x_m)$ in Eq.~\eqref{eq:encodingunitary} by $H_{\text{toy}}(\lambda x_m)$ and by replacing $|\mathbb{Z}_2\rangle$ by $|\boldsymbol{0}\rangle$:
\begin{equation}
    \kappa_{\text{toy}}(x, x') = |\langle \boldsymbol{0} | U_{\text{toy}}^{\dagger}(\lambda x;T) U_{\text{toy}}(\lambda x';T)|\boldsymbol{0}\rangle|^2 \,.
    \label{eq:toykernel}
\end{equation}
 
This approximate toy model captures the oscillatory behavior of the Rydberg-blockaded dynamics but not its quantum thermalization. In particular, it does not account for the damping of the oscillations due to the Rydberg dynamics leaking from the small subspace of quantum many-body scars to the exponential-size Hilbert space defined by the Rydberg-blockade constraints (which is seen both in numerical studies of Eq.~\eqref{eq:PXP} and in experiments). Moreover, it does not account for the volume-law entanglement generation of the Rydberg-blockaded dynamics \cite{choi2019emergent}.\\

\subsection{Perturbative toy kernel expression and analytical mean and variance}
\label{toymodelanalyt}
We now derive an analytical closed-form expression for this toy kernel, in the perturbative regime $\lambda \ll 1$ and in the single-feature case ($M=1$, $\boldsymbol{x}\equiv x$, $\boldsymbol{x'}\equiv x'$), using the linear response framework commonly employed in theoretical studies of the Loschmidt echo~\cite{GPSZLoschEcho2006} (full calculation details are available in Appendix.~\ref{supp-subseq:singlefeature}).

The idea is that the toy kernel Eq.~\eqref{eq:toykernel} is essentially the expectation value squared of a ``noisy'' echo operator, 
\begin{equation}
    M_{xx'}(\lambda, T) = U_{\text{toy}}^{\dagger}(\lambda x;T) U_{\text{toy}}(\lambda x';T) \, .
\end{equation}
Taking its time-derivative, we find its time behavior to be, 
\begin{equation}
    M_{xx'}(\lambda, t) = \mathcal{T} \exp \left( -i \lambda (x'-x)\frac{\Omega_{\text{toy}}}{2} \int_0^{T} \tilde{V}(\lambda, t') dt'\right) \,,
    \label{eq:echooperator}
\end{equation}
where $ \mathcal{T}$ designates the time-ordering operator and $\tilde{V}(\lambda, t')$ is the encoding perturbation $V$ in the interaction picture.

We use the Born expansion truncated to 2nd order in $\lambda$ to simplify it further, leading to a perturbative expression for the toy kernel,
\begin{align}
    \kappa_{\text{toy}} (x,x') &=\label{eq:toykernel_bornexp}\\ 1 - \lambda^2 (x'-x)^2 &\left(\frac{\Omega_{\text{toy}}}{2}\right)^2 \int_0^T \!\!dt' \int_0^T \!\!dt'' C_{\lambda}(t', t'') + \mathcal{O}(\lambda^4) \,, \nonumber
\end{align}
with $C_{\lambda}(t', t'') = \langle \tilde{V}(\lambda, t') \tilde{V}(\lambda, t'') \rangle - \langle \tilde{V}(\lambda, t') \rangle \langle \tilde{V}(\lambda, t'') \rangle$ a 2-point time-correlation function of the encoding perturbation. The computation of this correlator can be reduced to single-qubit calculations, leading to the following perturbative closed-form expression for the single-feature toy kernel: 
\begin{align}
    \begin{split}
    &\kappa_{\text{toy}} (x,x') =\\ &\;\;1 - \tilde{\lambda}^2 (x'-x)^2  N \sin^4 \left( \frac{\tilde{\Omega}_{\text{toy}} T}{2} \right) + \mathcal{O}(\tilde{\lambda}^4) \,, \label{eq:toykernelscaling}
    \end{split}
\end{align}
where $\tilde{\lambda}=2\delta /\tilde{\Omega}_{\text{toy}}$ and $\tilde{\Omega}_{\text{toy}} = \Omega_{\text{toy}} \sqrt{1+(\lambda x')^2}$ is the generalized Rabi frequency (details of this derivation can be found in Appendix~\ref{supp-subseq:toykernel}).

From Eq.~\eqref{eq:toykernelscaling}, we see that, if $x=x'$ then $\kappa_{\text{toy}}=1$. If instead $x\neq x'$, then $\kappa_{\text{toy}}\leq 1$  and decreases with $N$ and $\lambda^2$. Moreover, the toy kernel value oscillates with a period $\sim T_{\text{rev}}$ analogous to the spin precession itself. For fixed $N$ and $\lambda$, the toy kernel value is highest in the vicinity of revivals, which compensates the decrease but, at revival times, the value becomes essentially constant and thus data-independent , i.e., $\kappa_{\text{toy}}(x,x')=1 \, \forall x,x' \in [0,1]$ if $T=n T_{\text{rev}}$ ($n=1,2,\dots$).

Finally, we average over a random dataset to derive the mean and variance of the perturbative toy kernel in Eq.~\eqref{eq:toykernelscaling}. Considering single-feature data points $x,x' \in \mathcal{X}$ i.i.d. according to an arbitrary probability distribution, and by linearity of the expectation value, we find the mean and variance at fixed time $T$ to be
\begin{align}
    1 - \mathbb{E} [\kappa_{\text{toy}}(x, x')] &\propto N \lambda^2 \label{eq:toykernelmean} \,,\\
    \text{Var}[\kappa_{\text{toy}}(x, x')]  &\propto N^2 \lambda^4 \,.
    \label{eq:toykernelvar}
\end{align}
Thus, the mean and variance behave polynomially in $N$ and $\lambda$ , which is a direct consequence of the toy kernel being linear in $N$ and quadratic in $\lambda$ in the perturbative regime (Eq.~\eqref{eq:toykernelscaling}). Therefore, the toy kernel does not exhibit exponential concentration; in fact, the variance is guaranteed to remain non-decreasing as system size grows. The full details of this part of the calculation are available in Appendix~\ref{supp-subseq:toymeanvar}. In the next section, we provide evidence that the predicted behavior holds for \textsc{RydKernel}.\\

\begin{figure}[!tbp]
\includegraphics[width=\textwidth]{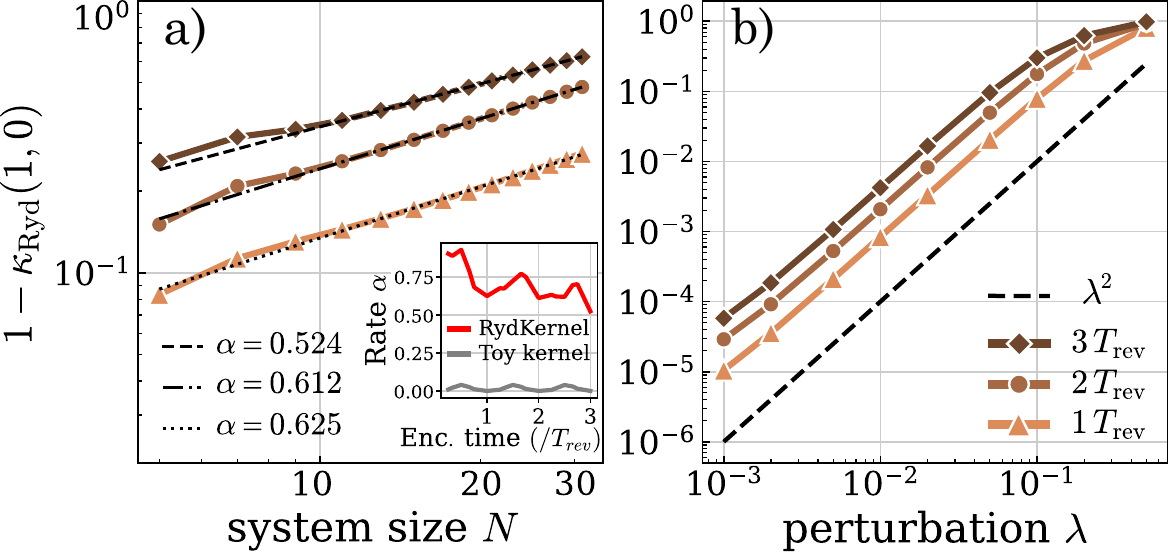} 
\caption{Scaling of $\kappa_{\text{Ryd}} (1,0)$ with respect to (a) system size $N$ and (b) encoding perturbation strength $\lambda$. We compare different encoding times $T=nT_{\text{rev}}$ ($n=1,2,3$). Simulations in (a) are done for $N\in [5,31]$ with $\lambda = 0.2$. Fits to $N^{\alpha}$ with rate $\alpha$ are performed on $N\geq 11$ to avoid finite-size effects (chi-square score $\lesssim 0.01$ for all fits). Inset: rates $\alpha$ for intermediate encoding times ($T\in  \mathopen]0,3T_{\text{rev}}\mathclose]$), in particular all integer multiples of $T_{\text{rev}}/3$ and $T_{\text{rev}}/4$. Simulations in (b) are done for $\lambda \in [0.001,0.5]$ with $N=31$. The black dashed line corresponding to $1-\kappa_{\text{toy}} \propto \lambda^2$ is shown as a guide to the eye. Intermediate times behave similarly and interpolate between $nT_{\text{rev}}$ curves. All simulations use TEBD with a maximum bond dimension of $\chi=580$.}
\label{fig:toykernel}
\end{figure}

\section{Main Results}
\label{resultsmain}
In this section, we first clarify the differences between \textsc{RydKernel} and the toy kernel, in particular at revival times. This motivates choosing precisely $T=n T_{\text{rev}}$ as encoding times for \textsc{RydKernel}. We then provide numerical evidence for the absence of exponential concentration in \textsc{RydKernel} in the corresponding regime of small encoding perturbation as predicted by the toy kernel. Furthermore, we show this EC-free behavior persists for strong encoding perturbations, up to $\lambda=1$, after which EC is recovered, delineating the boundaries of the EC-free phase of \textsc{RydKernel}. Finally, we check the generalization performance of the \textsc{RydKernel} on a standard  classical dataset.

\subsection{Comparison with the perturbative toy kernel}
\label{resultswithpert}
We showed in the previous section through analytics that the toy kernel value is expected to be decreasing with $N$ and $\lambda^2$, except at revival times, where the oscillating part, while alleviating this decrease, makes the kernel constant. This feature is a consequence of the perfect revivals of the toy model.

For \textsc{RydKernel}, however, this behaviour is not expected to hold, since the toy model does not capture the thermal features of the Rydberg setting, particularly the imperfect revivals in the latter. Thus, this data-independent behavior is not present in \textsc{RydKernel}. In fact, because of the high fidelity of the revivals and their weak sensitivity to the encoding perturbation, we choose precisely $T=n T_{\text{rev}}$ as encoding times for \textsc{RydKernel}. We expect this choice prevents concentration caused by measuring a global observable (the fidelity) to evaluate the kernel~\cite{thanasilp2024exponential}. 

To confirm this intuition, we compare the behavior of \textsc{RydKernel} to the toy kernel, with respect to changes in the system size $N$ and perturbation strength $\lambda$, at different encoding times fixed to integer multiples of the revival time $T_{\text{rev}}$. To do so, we fix $M=1$ and $x-x'=1$ and numerically simulate \textsc{RydKernel} in Eq.~\eqref{eq:rydbergkernel} in the ideal limit of the Rydberg blockade, where all interactions beyond nearest-neighbours (NN) in the Hamiltonian of Eq.~\eqref{eq:rydberg} are neglected. In all data presented in the paper, we use time-evolving block decimation (TEBD) simulations with parameters $\Omega=2$ and $V_{i, i+1}=4.4\Omega$. 

The results of our simulations are presented in Fig.~\ref{fig:toykernel}. Whereas the toy kernel is predicted to be constant and equal to $1$ at revival times, the infidelity of our kernel $1-\kappa_{\text{Ryd}}$ is instead found to be growing as $\sim N^{\alpha}$ with a rate $\alpha \sim \mathcal{O}(1)$ (Fig.~\ref{fig:toykernel}(a) and inset). The upward shift of infidelity as we go higher in the number of revivals is attributed to the imperfect oscillatory dynamics, though the growth rate $\alpha$ decreases with increased encoding time. 

Additionally, the dependence in the encoding perturbation strength $\lambda$ is quadratic (see dashed line in Fig.~\ref{fig:toykernel}(b), which follows the predictions of the toy kernel even at revivals, as long as one stays in the perturbative regime. The observed consistency with Eqs.~\eqref{eq:toykernelmean} and ~\eqref{eq:toykernelvar} confirms that \textsc{RydKernel} remains data-dependent at the revivals, and it is at these points that it encounters the slowest decrease in infidelity.

\subsection{Absence of exponential concentration}
\label{ECabsence_old}

\subsubsection{Perturbative regime}

\begin{figure}[!tbp]
\includegraphics[width=\textwidth]{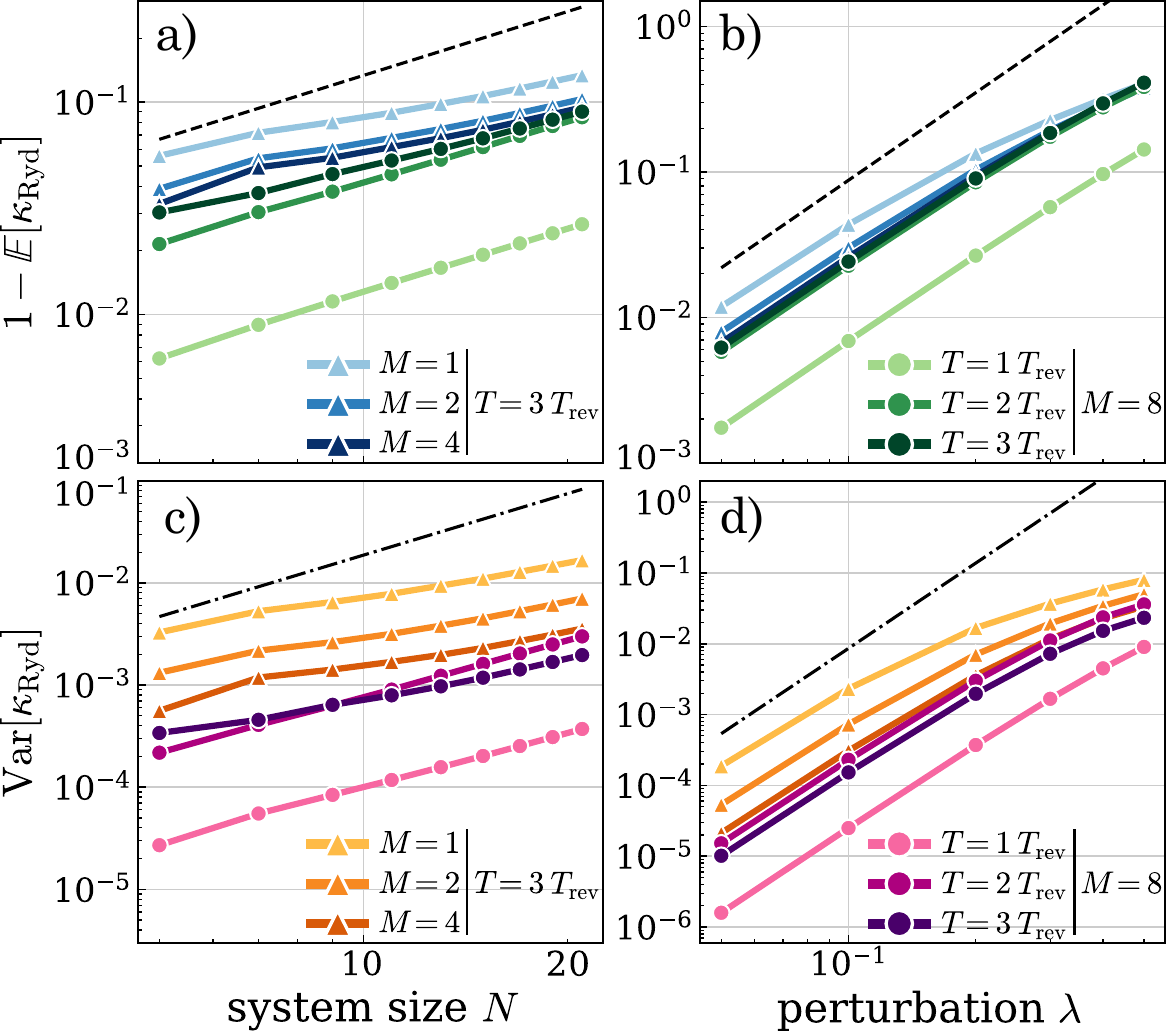} 
\caption{Moments of $\kappa_{\text{Ryd}}$ on a random dataset $\mathcal{X}$, with $|\mathcal{X}| = 20$ and $M$ features. Data points $0 \leq x_m \leq 1$ are chosen according to a uniform distribution with $\mu=0.5$ and $\sigma^2=1/15$. Mean and variance of the kernel off-diagonal elements are plotted against system size $N$ (a,c) and encoding perturbation strength $\lambda$ (b,d) for varied combinations of encoding times $T$ and number of features $M$. Simulations in (a,c) are done for $N\in [5,21]$ with $\lambda = 0.2$. Simulations in (b,d) are done for $\lambda \in [0.05,0.5]$ with $N=21$. Black dashed and dashed-dotted lines correspond to the scalings from Eqs.~\eqref{eq:toykernelmean} and \eqref{eq:toykernelvar}, respectively.}
\label{fig:compound3}
\end{figure}

We now investigate \textsc{RydKernel}'s concentration behavior for small encoding perturbation ($\lambda=0.2$). To do so we numerically compute the mean and variance of $\kappa_{\text{Ryd}}$ on random datasets $\mathcal{X}$ of real vectors $\boldsymbol{x}$ ($0 \leq x_m \leq 1$) with $M$ features and with $\left| \mathcal{X} \right| = 20$. 
Our results are shown in Fig.~\ref{fig:compound3}, where dashed and dashed-dotted lines correspond to the scaling forms of Eqs.~\eqref{eq:toykernelmean} and \eqref{eq:toykernelvar}.
The mean of the kernel (Fig.~\ref{fig:compound3}a) remains close to 1, with a decay rate that converges to the prediction of Eq.~\eqref{eq:toykernelmean} as $M$ increases.
Increasing the encoding time $T$ has little effect on the rate, while the magnitude of $\mathbb{E}[\kappa_{\text{Ryd}}]$ decreases slightly before stabilizing.

In Fig.~\ref{fig:compound3}c, the variance of the kernel exhibits at most quadratic scaling with system size, again aligning with the $\text{SU}(2)$ toy kernel. Most importantly, a positive rate indicates lack of concentration. While increasing $M$ leads to a decrease in the variance magnitude, the scaling remains unchanged.  In contrast, increasing $T$ results in a modest decrease in the scaling rate and a corresponding increase in variance magnitude, which eventually stabilizes. 
We also study the mean (Fig.~\ref{fig:compound3}b) and variance (Fig.~\ref{fig:compound3}d) of $\kappa_{\text{Ryd}}$ with respect to the encoding strength $\lambda$. Both follow quadratic and quartic scaling, respectively, as predicted by Eqs.~\eqref{eq:toykernelmean} and ~\eqref{eq:toykernelvar}. This holds irrespective of number of features or encoding time, as long as one remains within the regime of small perturbation.

The excellent agreement between the numerical results and the analytical predictions constitutes strong evidence that \textsc{RydKernel} is indeed free of exponential concentration in this regime.

\subsubsection{Beyond the perturbative regime\label{resultsbeyondpert_new}}

We extend our investigation of the EC-free behavior of \textsc{RydKernel} to the large encoding perturbation regime ($\lambda >0.2$).

\begin{figure}[!tbp]
\includegraphics[width=\textwidth]{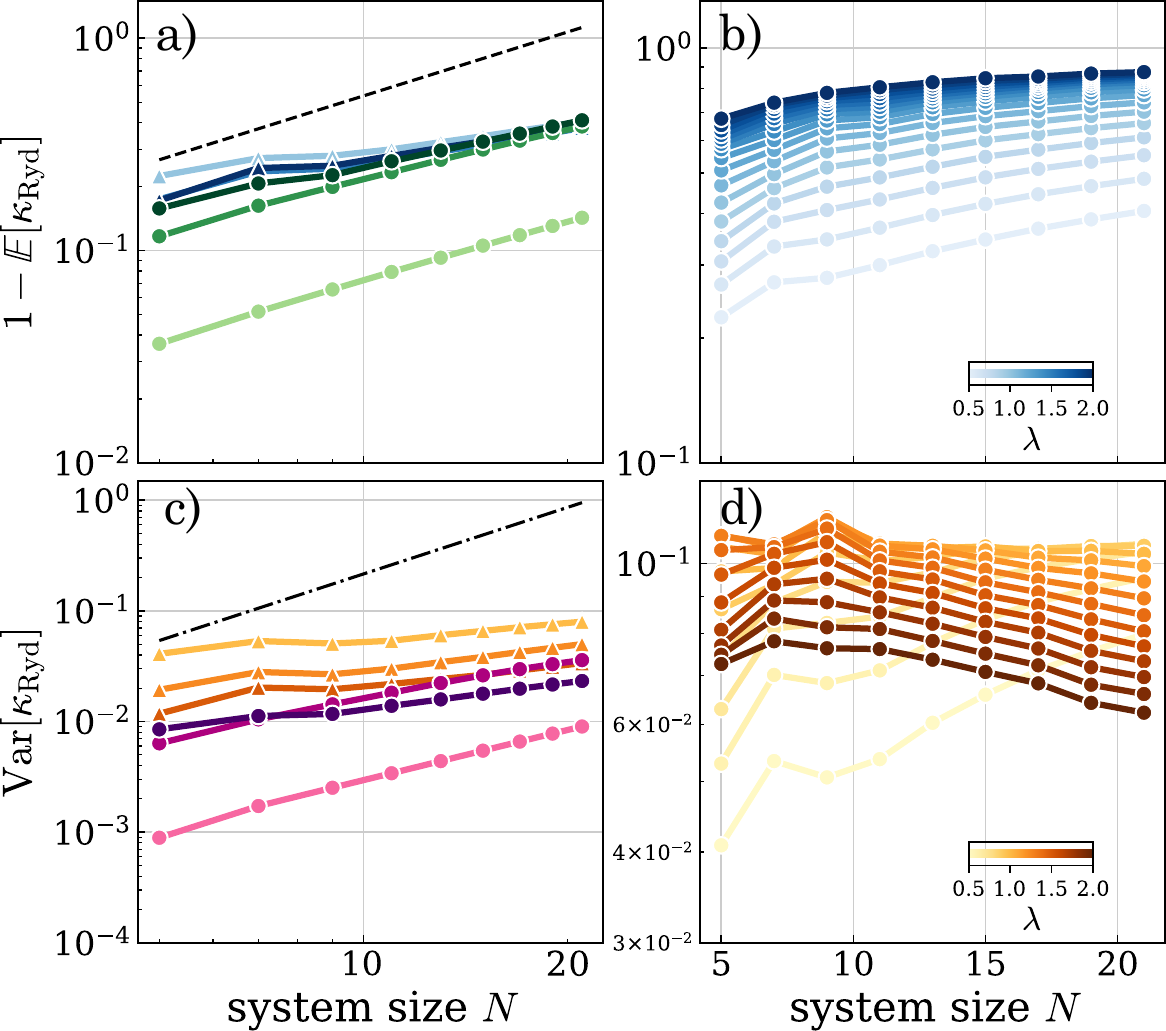} 
\caption{Moments of $\kappa_{\text{Ryd}}$ beyond the perturbative regime (random dataset $\mathcal{X}$, with $|\mathcal{X}| = 20$ and $M$ features). (a,c) Mean and variance against system size $N$ with $\lambda = 0.5$ (otherwise parameters and legend are identical to Fig. \ref{fig:compound3}). (b,d) Mean and variance against system size $N$ in log-scale for increasing values of the encoding perturbation magnitude $\lambda \in[0.5,2.0]$ with $0.1$ increments ($T = 3 T_{\text{rev}}$ and $M=1$). Darker colors corresponds to higher values of $\lambda$.}
\label{fig:compound4}
\end{figure}

We first perform in Fig. \ref{fig:compound4}(a,c) the same system size scaling analysis of the mean and variance on random datasets as in Fig. \ref{fig:compound3}(a,c) but with $\lambda=0.5$. Our results show very similar behaviors. The mean follows the same decay rate with a slightly lower magnitude (less than an order of magnitude). We notice that results for different number of features $M$ are also more clustered. The variance of the kernel is more than an order of magnitude higher than in the $\lambda=0.2$ case with still a positive but slightly lower scaling rate, indicating a persistent lack of concentration.

To identify the boundaries of \textsc{RydKernel}'s EC-free phase, we plot in Fig. \ref{fig:compound3}(b,d) the mean and the variance on the same random datasets for $M=1$ for increasing perturbation magnitudes, up to $\lambda=2.0$. We observe that the mean consistently lowers in magnitude as $\lambda$ is increased, towards values close to $0$. Most importantly, the scaling rate of the variance transitions from positive to negative around $\lambda \simeq 1.0$. Beyond this threshold, the variance of the kernel exhibits a clear exponentially decaying behavior, a signature that EC is retrieved. 

In summary, our numerical simulations confirm that EC-free behavior of the \textsc{RydKernel} persists even under relatively stronger encoding perturbations $(\lambda >0.2)$, and also identify a critical encoding strength $\lambda_c \simeq 1.0$ beyond which exponential concentration is recovered.

It is worth noting that the critical encoding strength $\lambda_c$ 
identified here is not a universal threshold, but rather specific to 
the dataset and encoding configuration used in 
Figs.~\ref{fig:compound3} and \ref{fig:compound4}. In the 
encoding scheme employed, the features $\{x_m\}_{m=1,\dots,M}$ of a data point $\boldsymbol{x}$ are encoded in successive layers of evolution under the Rydberg Hamiltonian of Eq.~\eqref{eq:rydberg}, each of duration $T/M$, with detuning strength $\lambda x_m$ (see Eq.~\eqref{eq:encodingunitary}). Consequently, for fixed $T$ and $\lambda$, increasing $M$ reduces the perturbative weight per layer, while increasing $\lambda$ or $T$ drives the system further from the perturbative regime. The threshold $\lambda_c$ is therefore expected to acquire a dependence on $M$, $T$, and the statistical distribution of the dataset.

\section{Hardness of classical simulations\label{hardnessofCS}}

\begin{figure*}[!tbp]
\includegraphics[width=\textwidth]{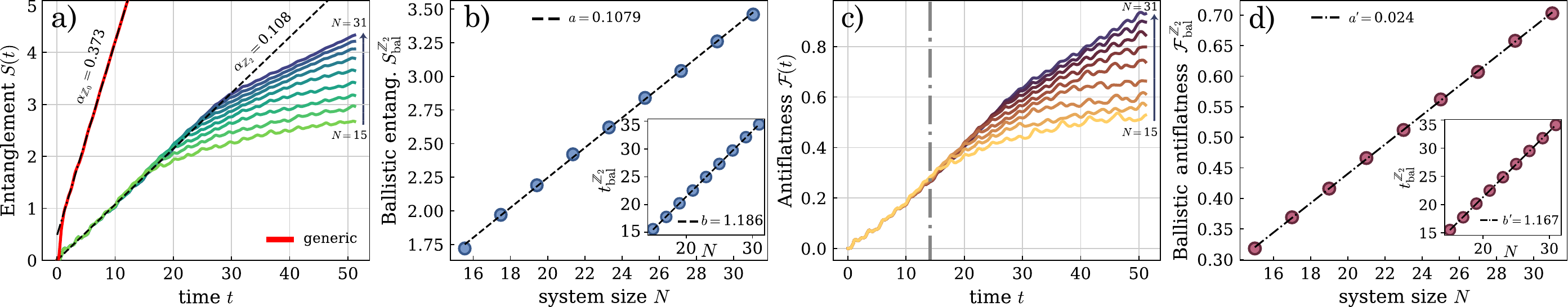} 
\caption{(a) Initial ballistic growth of mid-chain entanglement entropy $S(t)$ for the $\ket{\mathbb{Z}_2}$-Rydberg-blockaded dynamics ($\lambda=0$). (b) Scaling of entanglement entropy at the end of the ballistic regime (dashed line: linear fit $aN$). Inset: scaling of the duration of the entropy ballistic regime (dashed line: linear fit $bN$). (c) Time evolution of mid-chain anti-flatness $\mathcal{F}(t)$ for the $\ket{\mathbb{Z}_2}$-Rydberg-blockaded dynamics ($\lambda=0$). Gray dashed line indicates time $T=3.0 T_{\text{rev}}$. (d) Scaling of the anti-flatness at the end of the initial ballistic phase (dashed line: linear fit $a'N$). Inset: scaling of the duration of the anti-flatness ballistic phase (dashed line: linear fit $b'N$). TEBD simulations with a maximum bond dimension of $\chi=3070$.}
\label{fig:compound5}
\end{figure*}

Next, we quantify the hardness of classically simulating the proposed QKM, i.e., criterion (2) of the introduction, which boils down to characterizing the minimum cost for a classical computer to evaluate the fidelity $|\langle \psi(\boldsymbol{x},t) | \psi(\boldsymbol{x'},t)\rangle|^2$ between two highly entangled states resulting from the Rydberg-blockaded dynamics of the $|\mathbb{Z}_{2} \rangle$ state. 

No universal quantifier of hardness of classical simulation exists for quantum many-body dynamics. The landscape of computational frameworks is divided — each method is tailored to specific physical regimes and limited by different computational resources. Among the state-of-the art classical methods to simulate quantum many-body dynamics, tensor networks and stabilizer-based methods like Pauli-propagation stand out due to the high degree of control on computational errors. The former is limited by the amount of entanglement building up in the system, while the bottleneck of the latter is the amount of nonstabilizerness, also called magic. We discuss both resources in the context of the classical simulation of \textsc{RydKernel}.

\subsection{Entanglement growth}
 \label{entngrowth}

Tensor networks based on matrix product states (MPS) are among the best classical numerical methods for the simulation of one-dimensional quantum many-body dynamics.

The time-dependent variational principle (TDVP), for instance, allows for the evaluation of local observables at large system sizes. However, it does not have guaranteed accuracy at times beyond $\mathcal{O}(1)$ for global observables like the fidelity kernel \cite{Ho2019, turner2021correspondence,Sander2025}.

For finite system sizes, TEBD (and refined variations of it) is the leading tensor network method for simulating the behavior of global observables at long evolution times \cite{lin2020slow,Kerschbaumer2025,Shaw2024,Ljubotina2023}.
TEBD is based on matrix-product-state (MPS) methods, in which the precision of the classical simulation is limited by the largest accessible MPS bond dimension $\chi$. This, in turn, sets the maximum entanglement entropy $S_{\max} \sim \ln \chi$. 

The Rydberg-blockaded dynamics arising from the $|\mathbb{Z}_{2} \rangle$ initial state exhibits an early-time ballistic growth of the half-system entanglement entropy, $S(t)= a_{\mathbb{Z}_{2}} t$ (Fig. \ref{fig:compound5}a), with a system-size independent growth rate $a_{\mathbb{Z}_{2}}$ \cite{turner2018weak}. 
This growth is slower than the case of purely generic dynamics arising from the $|\mathbb{Z}_{0}\rangle = |0 \rangle^{\otimes N}$ initial state ($a_{\mathbb{Z}_{2}}\approx a_{\mathbb{Z}_{0}}/3.42$).  
Furthermore, as shown in Fig. \ref{fig:compound5}b, the short duration of this ballistic regime is proportional to system size ($ t_{\text{bal}}^{\mathbb{Z}_{2}} \approx 1.7 N/\Omega$), after which volume-law entanglement  is reached ($S_{\max} \sim N \log d$, with $d$ the local effective dimension). 
Simulations at long evolution times $t \geq t_{\text{bal}}^{\mathbb{Z}_{2}}$ thus require a bond dimension $\chi \sim d^N$.
Hence, given $\chi$, there exists a finite size $N_{\chi}$ and an associated finite time $t_{\chi}$ beyond which classical simulation is out of reach. Crucially, we find that this behavior remains present within the whole EC-free phase, $\lambda \in [0,1]$, and that the growth rate increases for higher $\lambda$ (see details in Appendix.~\ref{supp-subseq:tebdsim}).

A rough estimate of the memory required to represent the MPS, scaling as $\mathcal{O}(N \chi^2)$, of an $N=45$ qubit system (realizable experimentally in NAQCs) for the whole duration of the ballistic regime (up to $t_{\text{bal}}^{\mathbb{Z}_{2}} \approx 8 T_{\text{rev}}$) in double precision gives $\sim 1 \text{TB}$ ($\chi = 28643$). 
Moreover, kernel evaluation requires performing $\mathcal{O}(|\mathcal{X}|^2)$ of these TEBD simulations, each taking a time scaling as $\mathcal{O}(N \chi^3)$. 
Together, these scaling arguments tell us that, even for chains of moderate length, classical simulation of \textsc{RydKernel} becomes infeasible, fulfilling criterion (2). We note that a 2D implementation of our kernel based on the exact same physics, as has been observed experimentally~\cite{bluvstein2021controlling}, would be further beyond classical simulation.


\subsection{Magic growth}
\label{magicgrowth}
Probing entanglement growth is not enough to assess the hardness of classically simulating quantum dynamics. Clifford unitary dynamics are capable of generating highly entangled states, called stabilizer states, that can in fact be efficiently classically simulated (with $\mathcal{O}(N^2)$ memory space).

Sparse Pauli dynamics methods can be used to evaluate expectation of evolved operators by approximating their evolution in the Pauli operator basis \cite{Rall2019,Rudolph2025}. They are efficient when the number of Pauli strings to keep track of scales reasonably with the size of the system, or, when it is possible to get a faithful estimate of the target quantity by only keeping track of a reasonable number of strings while truncating others. This is precisely the case in Clifford dynamics, since Clifford operations map each Pauli string to only one other Pauli string, keeping the computational complexity low.

However, this kind of dynamics is not universal. The addition of one single-qubit non-Clifford gate, like the T gate, is enough to drive the dynamics outside of the stabilizer set and make it hard to simulate classically. While circuits with $\log(N)$ T gates can still be simulated in polynomial time, the runtime is in general exponential in the number of T gates. Concretely, this is because this non-Clifford resource, often called non-stabilizerness or magic, allows for branching dynamics, mapping each Pauli string to many others. The output state of such universal dynamics is in general a complex distribution over an exponential number of Pauli strings.

The amount of magic in the system at time $t$ of the evolution can be measured by the stabilizer 2-R\'{e}nyi entropy (SRE${_2}$) $\mathcal{M}_2 (\ket{\Psi(t)})$ of the state's Pauli distribution \cite{Leone-SRE2022-PhysRevLett.128.050402,Leone-stabilizerentropy-monotone-PhysRevA.110.L040403}. It quantifies how well-spread the distribution is and gives an estimate of the number of T gates. 

\textsc{RydKernel} as expressed in Eq.\eqref{eq:rydbergkernel} can be seen as the expectation value over the initial state $\ket{\mathds{Z}_{2}}$ of the nonlocal (pure) density operator $\dyad{\Psi_{\boldsymbol{x},\boldsymbol{x'}}(t)}{\Psi_{\boldsymbol{x},\boldsymbol{x'}}(t)}$ resulting from a parameterized Rydberg-blockaded dynamics, where $\ket{\Psi_{\boldsymbol{x},\boldsymbol{x'}}(t)}= U_{\text{Ryd}}^{\dagger}(\lambda\boldsymbol{x}, T) U_{\text{Ryd}}(\lambda \boldsymbol{x'}, T) \ket{\mathds{Z}_{2}}$.


In \cite{Smith2025}, $\text{SRE}_2$ generated along the effective PXP dynamics from the $\ket{\mathds{Z}_{2}}$ state was shown to grow slowly with a distinctive oscillating pattern, without ever going back to zero. On the other hand, magic created during precessing spin dynamics was shown to come back to zero periodically. Unfortunately, directly computing $\text{SRE}_2$ is numerically costly for entangled states with bond dimension $\chi \gtrsim 8$, which impedes its characterization at times longer than $\sim 1T_{\text{rev}}$ in the Rydberg dynamics.

What distinguishes genuine complex many-body nonstabilizerness from the one present in a product of single-qubit magic states is the presence of non-local magic. Non-local magic is the part of magic that lives in the correlations between subsystems and cannot be removed by local operations. Thus, magic is lower-bounded by non-local magic.

Most importantly, it was recently shown that non-local magic can be probed by looking at a higher-moment feature of the entanglement spectrum called anti-flatness \cite{Tirrito_Antiflatness_PhysRevA.109.L040401,Cao2024_graviBR_magic_z3vr-w5c5,nonlocalstabilizerness-PhysRevA.111.052443}. This quantity, defined for subsystem density matrix $\rho_A$ of an arbitrary bipartition $AB$ as
\begin{equation}
    \mathcal{F}(\rho_A) = \frac{\text{Tr}(\rho_A^3) - \text{Tr}^2(\rho_A^2)}{\text{Tr}^2(\rho_A^2)} \; ,
\end{equation}
lower-bounds non-local magic, and thus magic itself, and is much easier to compute than non-local SRE. We therefore use it as a probe of non-stabilizerness in \textsc{RydKernel}.

In Fig.~\ref{fig:compound5}(c) we show that the realistic Rydberg-blockaded dynamics underlying our kernel exhibits increasing non-zero anti-flatness for $t>0$ up to very long times ($\sim 11 T_{\text{rev}}$). This holds for any value of the perturbation within the EC-free phase (see Appendix.~\ref{supp seq: NL magic}).

Anti-flatness initially grows at a rate that is independent of system size, rapidly approaching the Haar-random-state prediction of 0.25 \cite{Odavic2025}. This occurs even for moderate system sizes ($N\gtrsim 15$ at $T=3T_{\text{rev}}$), remaining above this value for all investigated times and system sizes thereafter. This is to be contrasted with the behavior in random unitary circuits which shows a peak at times  $t \lesssim O(\log N)$, where most magic is developed (see Appendix~\ref{supp seq: NL magic}), followed by a saturation at the finite Haar random value \cite{TurkeshiRQC2025-Natcomm}, or in non-abelian lattice gauge theories dynamics, which exhibit a peak at short times and a lower saturation value \cite{Ebner2025}. We conjecture that, in the case of $\ket{\mathds{Z}_{2}}$-Rydberg-blockaded dynamics, this intermediate phase where magic peaks, dubbed ``magic barrier", may extend for extremely long times (which matches the timescales for the saturation of entanglement entropy). These results offer, to the best of our knowledge, a novel perspective on the complexity of thermalization dynamics in weak-ergodicity breaking systems, complementing traditional entanglement entropy, and at times extending significantly beyond those in previous studies.

Crucially, as shown in Fig. \ref{fig:compound5}(d), we find that the duration of the initial growth is proportional to system size, as well as the corresponding anti-flatness, $\mathcal{F}\sim N$. This behavior is similar to entanglement entropy. Since anti-flatness rigorously lower-bounds magic (and non-local magic), this means magic itself must scale with system size even at moderate times, $\mathcal{M}_2 \sim N$. Consequently, classically simulating evolution times linear in $N$ with Pauli propagation techniques requires resources exponential in the system size.

This means the $\ket{\mathds{Z}_{2}}$ Rydberg-blockaded dynamics is highly complex, and we expect it to be resistant to numerical methods involving Clifford disentangling techniques such as Clifford-enhanced TEBD or TDVP~\cite{qian2024cliffordcircuitsaugmentedtimedependent,Qian2024-DMRGClifford-PhysRevLett.133.190402,masotllima2026limitsclifforddisentanglingtensor,fux2025disentanglingmagicstatesclassically}, or involving Monte-Carlo sampling of Pauli strings \cite{Sinibaldi2025}.

\section{Tasks of practical utility: Classification \label{MLperf}}

\begin{figure}[!tbp]
\includegraphics[width=0.98\textwidth]{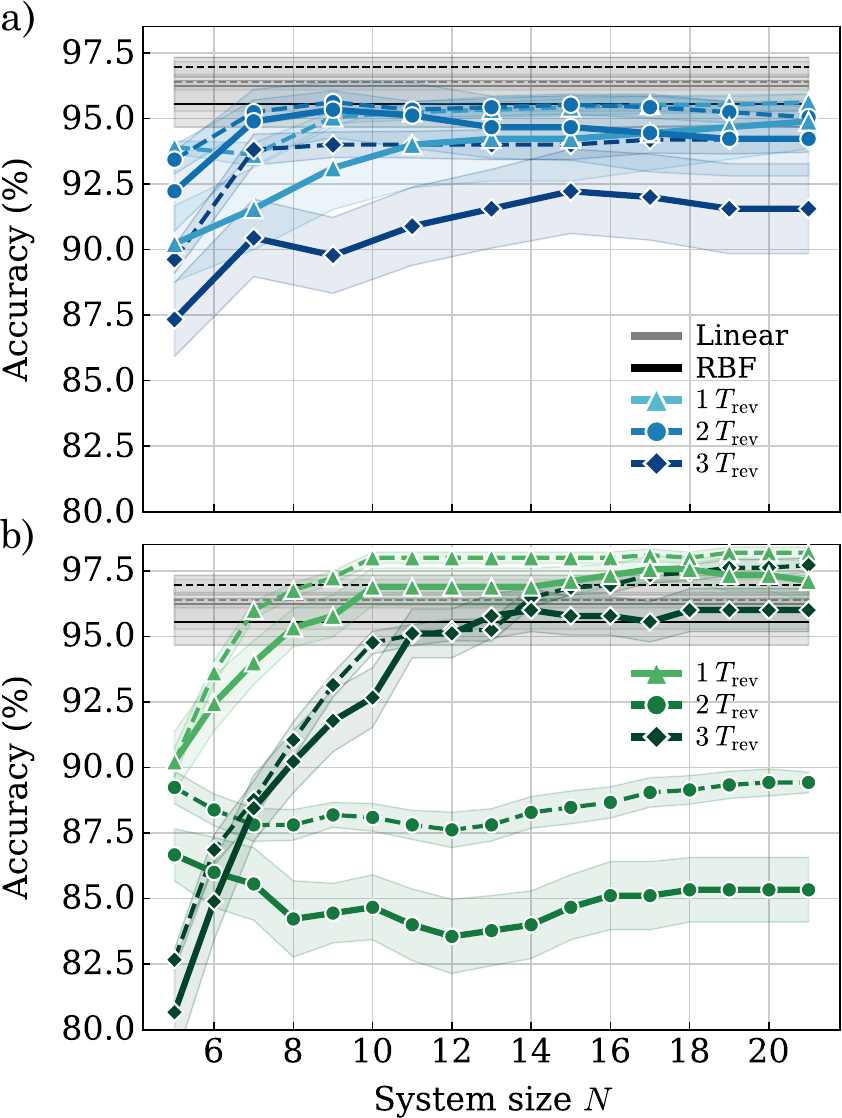} 
\caption{Classification performance of \textsc{RydKernel} (a) and the toy kernel (b) on the IRIS dataset at different encoding times $T=nT_{\text{rev}}$ ($n=1,2,3$) compared against the classical Linear and RBF kernels. Solid (dashed) lines correspond to test (resp. train) accuracies. We use $70\%$ of the dataset for training and $30\%$ for testing. Encoding strength of both $\kappa_{\text{Ryd}}$ and $\kappa_{\text{toy}}$ is set to $\lambda = 2.5/\sqrt{N}$.}
\label{fig:MLTaskIRIS}
\end{figure}

As a first benchmark of \textsc{RydKernel}, we evaluate its performance on a supervised classification task using the Iris dataset ($|\mathcal{X}| = 150$, $M = 4$, $|\mathcal{Y}| = 3$), comparing against radial basis function (RBF), linear, and the toy kernels (see Appendix \ref{supp-seq:mltasks} for details and definitions). A support vector machine is trained using \textsc{scikit-learn} with precomputed kernel values, and all reported accuracies are obtained via 10-fold cross-validation, with shaded regions indicating the variance across folds. Figures~\ref{fig:MLTaskIRIS}(a,b) show the training and test accuracies of \textsc{RydKernel} and the toy kernel as a function of $N$, for encoding times $T_{\text{rev}}$ to $3T_{\text{rev}}$, with $\lambda = 2.5/\sqrt{N}$. This scaling choice is motivated by Eqs.~\eqref{eq:toykernelmean}--\eqref{eq:toykernelvar} as it ensures that the mean and variance are scale-independent. Furthermore, this scaling keeps \textsc{RydKernel} outside the exponential concentration regime and yields steadily improving accuracy with $N$. At $T = 2T_{\text{rev}}$, \textsc{RydKernel} exceeds $95\%$ test accuracy near $N = 9$, approaching RBF performance — and notably, longer encoding times correspond to dynamics that are harder to simulate classically, so peak performance is reached precisely where classical emulation becomes costly. The toy kernel performs well near odd multiples of $T_{\text{rev}}$, even surpassing the linear kernel baseline at $T = T_{\text{rev}}$, but degrades substantially at even multiples (cf. $2T_{\text{rev}}$ ), where \textsc{RydKernel} clearly outperforms it. This even/odd structure is corroborated across a broader sweep of encoding times and strengths in Figs.~\ref{fig:rydk_toyk_phasediag15} and~\ref{fig:toyk_phasediag1000} in Appendix~\ref{supp-seq:irisclassif}.

We emphasize that these benchmarks do not demonstrate quantum advantage or claim superiority over classical kernel methods; rather, they serve as a proof-of-concept that \textsc{RydKernel} is a viable and trainable model on standard machine learning tasks. The central contribution of this work is the introduction of a class of physically-motivated quantum kernels rooted in Rydberg-physics that naturally circumvents exponential concentration — a property analytically established in Eqs.~\eqref{eq:toykernelmean}--\eqref{eq:toykernelvar} and numerically confirmed. Crucially, \textsc{RydKernel} remains well-behaved in the perturbative regime as system size grows not through post-hoc regularization, but as a direct consequence of the underlying physics. While the optimal data encoding strategy remains an open question, and different tasks may favor different encodings, the favorable anti-concentration properties and physical structure of \textsc{RydKernel} provide compelling motivation for future investigation into other ML tasks that might offer more natural settings for rigorous computational advantage.

\section{EXPERIMENTAL IMPLEMENTATION}
\label{expimplmntation}
Finally, we address criterion (3) and show that \textsc{RydKernel} is directly compatible with existing NAQCs.
These systems can be prepared in an initial $|\mathbb{Z}_2\rangle = U_{\mathbb{Z}_2}|\boldsymbol{0}\rangle$ state using either purely analog methods~\cite{bernien2017probing} or schemes involving local adressing of the qubits \cite{zhang2024analog, Bluvstein2022,LocalAddressing-66s8-jj18}. The latter involves a phase mask that imparts a set spatial modulation to the laser's detuning \cite{deoliveira2025quantumwireapproachweighted}.
By construction, the Hamiltonian of Eq.~\eqref{eq:rydberg} and the unitary evolution of Eq.~\eqref{eq:encodingunitary} are natively implemented in this platform. 
With these two ingredients in place, the simplest implementation of \textsc{RydKernel} is through a Loschmidt echo, a widely used protocol in the study of quantum many-body dynamics ~\cite{GPSZLoschEcho2006}. 
This requires one to reverse the sign of the Rydberg Hamiltonian to implement $U_{\text{Ryd}}^{\dagger}(\boldsymbol{x}, T)$ as a ``backward" evolution following the ``forward" part, $U_{\text{Ryd}}(\boldsymbol{x'}, T)$. 
In the PXP limit $H_{\text{PXP}}$, where next-nearest neighbor interactions are negligible, one observes that $Z H_{\text{PXP}} Z = - H_{\text{PXP}}$~\cite{Xiang2024,Liang2024}. 
Therefore, inserting a global $Z=\prod_i^N\sigma_i^z$ pulse between two forward evolutions under $H_{\text{PXP}}$ results in the wanted Lochsmidt echo form (see Appendix.~\ref{supp-subseq:leprotocol} for details).
This approach \textit{does not} reverse the sign of next-nearest neighbor interaction terms, which can negatively impact performance. 
This can be mitigated via perturbative Rydberg gadgets that reduce the relative magnitude of these terms at the cost of a linear overhead in the number of atoms~\cite{pichler2018computational, nguyen2023quantum, lanthaler2024quantum, Bombieri2025}. Additionally, it has been observed that the underlying Rydberg-blockaded dynamics is robust to finite-temperature effects~\cite{Desaules2023} and disorder~\cite{MondragonShem2021}, making experimental implementation viable in the near term.

Alternatively, the kernel can be evaluated using a SWAP-test protocol, which avoids the need for implementing time-reversed evolution. This has already been experimentally demonstrated in~\cite{Bluvstein2022} and require transferring the Rydberg state $\ket{r}$ to a hyperfine state $\ket{h}$, where digital operations can be performed, thereby introducing an additional source of infidelity. We provide more details on this in the implementation section of Appendix.~\ref{supp-seq:implementation}. 

While \textsc{RydKernel} exploits natural properties of analog neutral-atom platforms, the framework by itself is not platform-specific. In fact, the effective PXP dynamics underlying this has been successfully implemented on digital superconducting quantum hardware~\cite{Ladecola2022-DigPXP, ScarDigital-Desaules20} (as well as other scar-hosting models), a setting in which it can be made fault-tolerant.

\section{DISCUSSION AND CONCLUSION}
We underline three key points related to the performance and design of our QKM. 

Firstly, the toy kernel constitutes a viable machine learning kernel, as evidenced by the learning and generalization curves. However, owing to its integrable and non-entangling nature, the toy kernel remains efficiently classically simulable offering no quantum advantage. It can thus be interpreted as a quantum-inspired classical baseline. Recent works have highlighted precisely this tension between non-exponential concentration and classical simulability \cite{cerezo2023, diaz2023}. Here we make this distinction explicit: while both kernels avoid exponential concentration, \textsc{RydKernel} is not classically simulable, unlike the toy model, establishing a concrete separation in their potential for quantum advantage.

Secondly, another crucial feature of $\kappa_{\text{Ryd}}$ is that, while it hosts an approximate $\text{SU}(2)$ algebra, it also accesses an exponentially large sector of the Hilbert space, rendering classical simulation hard.
We suspect that this peculiar feature of the dynamics, related to Hilbert space fragmentation~\cite{Moudgalya_2022}, is responsible for the expressivity of the kernel. 

Thirdly, it was pointed out in Ref. ~\cite{thanasilp2024exponential} that QKMs based on measurement of a global observable in an exponential Hilbert space are prone to EC. 
Here, our choice of encoding time (targeting high-fidelity revivals) and data embedding (using a perturbation known to minimally impact revival fidelity~\cite{turner2018quantum}) likely plays a role in preventing EC. However, persistence of EC-freeness beyond the perturbative regime, where revivals gradually vanish while QMBS are preserved, suggest the latter are essential. We note that the approximate SU(2) group structure is also reminiscent of recent EC-free quantum kernel constructions \cite{henderson2025}. 

In summary, we introduced a QKM grounded in the Rydberg-blockade mechanism arising naturally in arrays of strongly interacting neutral atoms. 
We compared extensive numerical simulations with analytical results to demonstrate that \textsc{RydKernel} avoids EC by design, all the while being classically hard to simulate. 
This was established through entanglement and non-stabilizerness arguments, the latter offering a novel perspective on the thermalization complexity of weak ergodicity-breaking dynamics. 
We further demonstrated effective classification and generalization on the IRIS benchmark dataset and provided evidence for the realizability of \textsc{RydKernel} on current NAQC's.

Our results highlight the potential of harnessing unconventional quantum many-body dynamics for ML.
We believe it is worthwhile to gain deeper understanding of the Rydberg kernel. Studying the dynamics of different initial states in the Rydberg blockade can pinpoint the role of quantum many-body scars and fragmentation in kernel performance, with an eye towards accelerating ML tasks. Finding learning tasks and architectures where RydKernel demonstrates a genuine advantage over classical ML methods is another important avenue under investigation.
Furthermore, the question of hardware noise — and in particular the characterization of how different noise models native to NAQCs affect kernel concentration, trainability, and generalization - constitutes an important and natural goal for future work. Finally, studying \textsc{RydKernel} in 2D arrays, where revivals have also been observed experimentally~\cite{lin2020quantum,bluvstein2021controlling} and where classical simulations are further out of reach, is an exciting, yet challenging future direction.

\section{Acknowledgement}
The project was conceived by S.K. Theoretical results were proved by A.S., with inputs from V.D.T. and M.S. Numerical implementation was performed by M.S., and initial code prototyping was assisted by A.S., R.R. and S.D.K. Initial literature review was carried out by A.S., R.R, M.F., V.D.T. The manuscript was drafted by A.S. and M.S. and finalized by V.D.T. and S.K. The project was supervised by V.D.T. and S.K. We also thank J\'{e}r\'{e}mie Gince for critical reading of the manuscript. This research was financially supported by Prompt through its \textit{Soutien aux organismes de recherche} program, by the Natural Sciences and Engineering Research Council of Canada (NSERC) through an Alliance grant, and by Pasqal Canada through its contribution to the Institutional Research Chair in Quantum Artificial Intelligence at Université de Sherbrooke. AS acknowledges support from the Canada First Research Excellence Fund through an Institut Quantique Postdoctoral Fellowship. This work made use of compute resources by Calcul Québec and the Digital Research Alliance of Canada.

\bibliography{biblio}

\pagebreak


\textbf{Contents}
\begin{itemize}
    \setlength{\itemsep}{0pt}
    \item[\ref{supp-subseq:toykernel}] - Toy kernel analytics.
    \begin{itemize}
        \item[\ref{supp-subseq:singlefeature}] - Linear response behavior for a single feature. 
        \item[\ref{supp-subseq:toymeanvar}] - Mean and variance of the single-feature toy-kernel.
    \end{itemize}
    \item[\ref{supp-seq:classsim}] - Details on the classical simulation of \textsc{RydKernel}.
    \begin{itemize}
        \item[\ref{supp-subseq:tebdsim}] - TEBD simulations of one-dimensional \textsc{RydKernel}.
        \item[\ref{supp-subseq:hardness}] - Hardness of classical simulations.
    \end{itemize}
    \item[\ref{supp-seq:implementation}] - Details on the experimental implementation of \textsc{RydKernel} on analog neutral-atom quantum computer.
    \begin{itemize}
        \item[\ref{supp-subseq:naqc}] - Analog neutral-atom quantum computers (NAQC).
        \item[\ref{supp-subseq:leprotocol}] - \textsc{RydKernel} as a Loschmidt echo protocol.
        \item[\ref{supp-subseq:concretetimescales}] - Concrete timescales of a minimal implementation of the Loschmidt echo \textsc{RydKernel}.
    \end{itemize}
    \item[\ref{supp-seq:irisclassif}] - Additional classification results on the Iris dataset.
    \item[\ref{supp-seq:mltasks}] - Details on the classical machine learning methods. 
    \begin{itemize}
        \item[\ref{supp-subseq:svm}] - How the kernel is used in a support vector machine (SVM).
        \item[\ref{supp-subseq:classkernel}] - Classical kernel methods used in the paper.
    \end{itemize}
\end{itemize}
\vspace{1.5cm}

\appendix
\section{Toy kernel analytics \label{supp-subseq:toykernel}}
\subsection{Linear response theory of the toy kernel for a single feature \label{supp-subseq:singlefeature}}

In this section, we provide details about the analytical calculations leading to the perturbative expression for the single-feature toy kernel (Eq. \eqref{eq:toykernelscaling} in the main text). To make these calculations self contained, we rewrite here some parts of the calculation already mentioned in the main text. In this section, we use $y$ instead of $x'$ for visual clarity. Thus, for single-feature data, we set $\vec{x}\equiv x$, $\vec{y} \equiv y$, with $0 \leq x,y \leq 1$. The toy kernel is given in Eq. \eqref{eq:toykernel} where the unitary embedding $U_{\text{toy}}(\lambda x;T) = \exp\left( -i T H_{\text{toy}} (\lambda x) \right)$ is generated by the parametrized toy-model Hamiltonian given in Eq.\eqref{eq:toymodel}.  $T$ is the encoding time and $\lambda = \frac{2\delta}{\Omega_{\text{toy}}}$ is the magnitude of the detuning perturbation used to encode the data points.\\

The idea is that Eq.~\eqref{eq:toykernel} is essentially the expectation value squared of a ``noisy'' echo operator $M_{xy}(\lambda, T) = U_{\text{toy}}^{\dagger}(\lambda y;T) U_{\text{toy}}(\lambda x;T)$ over the initial state $|\boldsymbol{0}\rangle$. The time behavior of this operator is given by the time derivative
\begin{equation}
    \frac{d M_{xy}(\lambda, T)}{dt} = - i \lambda (x-y) \frac{\Omega_{\text{toy}}}{2} \tilde{V}(\lambda, T) M_{xy}(\lambda, T),
\end{equation}
giving the time-ordered expression Eq. \eqref{eq:echooperator} in the main text, with the encoding perturbation now represented in the interaction picture,
\begin{align}
    \tilde{V}(\lambda, t) &= U_{\text{toy}}^{\dagger}(\lambda y;t) \ V \ U_{\text{toy}}(\lambda y;t) \\ &= e^{i t \frac{\Omega_{\text{toy}}}{2} \left(S^{x} + \lambda y V \right)} \ V \ e^{-i t \frac{\Omega_{\text{toy}}}{2} \left(S^{x} + \lambda y V \right)} .
\end{align}
The next technical step is to compute the correlator in the 2nd order perturbative expression (in $\lambda$) of the toy kernel in Eq. \eqref{eq:toykernel_bornexp}, given by
\begin{equation}
    C_{\lambda}(t', t'') = \langle \tilde{V}(\lambda, t') \tilde{V}(\lambda, t'') \rangle - \langle \tilde{V}(\lambda, t') \rangle \langle \tilde{V}(\lambda, t'') \rangle \ .
\end{equation}
Since $S^x$ and $V$ are sums of single-site operators, we reduce the calculation of $\tilde{V}$ to a single-site calculation:
\begin{widetext}
\begin{align}
    \tilde{V}(\lambda, t) &= \exp\left[i t \frac{\Omega_{\text{toy}}}{2} \sum_{i}\left(  \sigma^{x}_{i} + \lambda y (-1)^{i} \sigma_{i}^{z} \right) \right] \left( \sum_{j}(-1)^{j} \sigma_{j}^{z} \right) \exp\left[-i t \frac{\Omega_{\text{toy}}}{2} \sum_{k}\left( \sigma^{x}_{k} + \lambda y (-1)^{k} \sigma_{k}^{z}\right) \right] \\ &= \sum_{j}(-1)^{j} \left[ \prod_{i} \exp\left[i t \frac{\Omega_{\text{toy}}}{2} \left( \sigma^{x}_{i} + \lambda y (-1)^{i} \sigma_{i}^{z} \right) \right] \left(  \sigma_{j}^{z} \right) \prod_{k} \exp\left[-i t \frac{\Omega_{\text{toy}}}{2} \left( \sigma^{x}_{k} + \lambda y (-1)^{k} \sigma_{k}^{z}\right) \right] \right] .
\end{align}
Because operators at site $i$ commute with those at site $i'\neq i$ and at site $j\neq i$, we get
\begin{equation}
    \tilde{V}(\lambda, t) = \sum_{j}(-1)^{j} \exp\left[i t \frac{\Omega_{\text{toy}}}{2} \left( \sigma^{x}_{j} + \lambda y (-1)^{j} \sigma_{j}^{z} \right) \right] \sigma_{j}^{z} \exp\left[-i t \frac{\Omega_{\text{toy}}}{2} \left( \sigma^{x}_{j} + \lambda y (-1)^{j} \sigma_{j}^{z}\right) \right] .
\end{equation}
Now, we use the general identity for Pauli matrices: $e^{ia(\hat{n} \cdot \vec{\sigma})} = \mathds{1} \cos(a) + i (\hat{n} \cdot \vec{\sigma}) \sin(a)$, with, in our case, $a=\pm \frac{\Omega_{\text{toy}} t}{2}\sqrt{1+(\lambda y)^2}$ (depending on the sign in the exponential), and $\hat{n} = \frac{1}{\sqrt{1+(\lambda y)^2}} \begin{pmatrix}
    1 \\ 0 \\ (-1)^j \lambda y
\end{pmatrix}$ a unit vector.
Setting $\tilde{\Omega}_{\text{toy}} = \Omega_{\text{toy}} \sqrt{1+(\lambda y)^2}$ for simplicity, we get
\begin{equation} 
    \tilde{V}(\lambda, t) = \sum\limits_{j}(-1)^{j} \left[ \mathds{1} \cos(\frac{\tilde{\Omega}_{\text{toy}} t}{2}) + i \frac{(\sigma_{j}^{x} + (-1)^j \lambda y \sigma_{j}^{z})}{\sqrt{1+(\lambda y)^2}} \sin(\frac{\tilde{\Omega}_{\text{toy}} t}{2})  \right] \sigma_{j}^{z} \left[ \mathds{1} \cos(\frac{\tilde{\Omega}_{\text{toy}} t}{2}) - i \frac{(\sigma_{j}^{x} + (-1)^j \lambda y \sigma_{j}^{z})}{\sqrt{1+(\lambda y)^2}} \sin(\frac{\tilde{\Omega}_{\text{toy}} t}{2}) \right]
\end{equation}
Expanding this expression leads to nine terms:
\begin{align} 
    \tilde{V}(\lambda, t) = \sum_{j}(-1)^{j} \Bigg[ &\sigma_{j}^{z} \cos^2(\frac{\tilde{\Omega}_{\text{toy}} t}{2}) - \frac{i}{\sqrt{1+(\lambda y)^2}} \underbrace{\sigma_{j}^{z} \sigma_{j}^{x}}_{i \sigma_{j}^{y}} \cos(\frac{\tilde{\Omega}_{\text{toy}} t}{2}) \sin(\frac{\tilde{\Omega}_{\text{toy}} t}{2}) - \frac{i (-1)^j \lambda y}{\sqrt{1+(\lambda y)^2}} \underbrace{\sigma_{j}^{z} \sigma_{j}^{z}}_{\mathds{1}} \cos(\frac{\tilde{\Omega}_{\text{toy}} t}{2}) \sin(\frac{\tilde{\Omega}_{\text{toy}} t}{2}) \nonumber \\ &+\frac{i}{\sqrt{1+(\lambda y)^2}} \underbrace{\sigma_{j}^{x} \sigma_{j}^{z}}_{-i \sigma_{j}^{y}} \cos(\frac{\tilde{\Omega}_{\text{toy}} t}{2}) \sin(\frac{\tilde{\Omega}_{\text{toy}} t}{2}) - \left( \frac{i}{\sqrt{1+(\lambda y)^2}} \right)^2 \underbrace{\sigma_{j}^{x} \sigma_{j}^{z} \sigma_{j}^{x}}_{-\sigma_j^z} \sin^2(\frac{\tilde{\Omega}_{\text{toy}} t}{2}) \nonumber \\ &- \left( \frac{i}{\sqrt{1+(\lambda y)^2}} \right)^2 (-1)^j \lambda y \underbrace{\sigma_{j}^{x} \sigma_{j}^{z} \sigma_{j}^{z}}_{\sigma_j^x} \sin^2(\frac{\tilde{\Omega}_{\text{toy}} t}{2}) +\frac{i(-1)^j \lambda y}{\sqrt{1+(\lambda y)^2}} \underbrace{\sigma_{j}^{z} \sigma_{j}^{z}}_{\mathds{1}} \cos(\frac{\tilde{\Omega}_{\text{toy}} t}{2}) \sin(\frac{\tilde{\Omega}_{\text{toy}} t}{2}) \nonumber \\ &- \left( \frac{i}{\sqrt{1+(\lambda y)^2}} \right)^2 (-1)^j \lambda y \underbrace{\sigma_{j}^{z} \sigma_{j}^{z} \sigma_{j}^{x}}_{\sigma_j^x} \sin^2(\frac{\tilde{\Omega}_{\text{toy}} t}{2}) - \left( \frac{i}{\sqrt{1+(\lambda y)^2}} \right)^2 (-1)^{2j} (\lambda y)^{2} \underbrace{\sigma_{j}^{z} \sigma_{j}^{z} \sigma_{j}^{z}}_{\sigma_j^z} \sin^2(\frac{\tilde{\Omega}_{\text{toy}} t}{2}) \Bigg] \,.
\end{align}
These are grouped Pauli-wise to get the final expression of the encoding perturbation in the interaction representation,
\begin{equation} 
    \tilde{V}(\lambda, t) = \sum_j (-1)^j \left[ \frac{(\lambda y)^2 + \cos(\tilde{\Omega}_{\text{toy}} t)}{1 + (\lambda y)^2} \, \sigma_j^z + \frac{\sin(\tilde{\Omega}_{\text{toy}} t)}{\sqrt{1 + (\lambda y)^2}} \, \sigma_j^y + \frac{(-1)^j \lambda y \big(1 - \cos(\tilde{\Omega}_{\text{toy}} t) \big)}{1 + (\lambda y)^2} \, \sigma_j^x\right] .
\end{equation}\\

The next step is to compute the correlator $C_{\lambda}(t', t'') = \langle \tilde{V}(\lambda, t') \tilde{V}(\lambda, t'') \rangle - \langle \tilde{V}(\lambda, t') \rangle \langle \tilde{V}(\lambda, t'') \rangle$ by taking expectation values with respect to the initial state $|\psi_0\rangle = |0\rangle^{\otimes N}$:
\begin{itemize}
    \item For the single $\tilde{V}$ expectation value, only $\sigma_j^z$ contributes, giving
    $$\langle \tilde{V}(\lambda, t')\rangle \langle \tilde{V}(\lambda, t'')\rangle  = \frac{1 - (-1)^N}{2} \frac{\left((\lambda y)^2 + \cos(\tilde{\Omega}_{\text{toy}} t')\right)\left((\lambda y)^2 + \cos(\tilde{\Omega}_{\text{toy}} t'')\right)}{(1 + (\lambda y)^2)^2}.$$
    \item For the connected part, only terms with a single $\sigma_j^x$ or a single $\sigma_j^y$ in $\tilde{V}(t') \tilde{V}(t'')$ do \underline{not} contribute, giving
    $$
    \!
    \begin{aligned}[t]
        \langle \tilde{V}(\lambda, t') \tilde{V}(\lambda, t'') \rangle &= N \frac{(\lambda y)^2}{(1+(\lambda y)^2)^2} \left(1-\cos(\tilde{\Omega}_{\text{toy}} t')\right) \left(1-\cos(\tilde{\Omega}_{\text{toy}} t'')\right) + N\frac{1}{1+(\lambda y)^2} \sin(\tilde{\Omega}_{\text{toy}} t') \sin(\tilde{\Omega}_{\text{toy}} t'') \\ & \qquad \qquad+ \frac{1 - (-1)^N}{2} \frac{\left((\lambda y)^2 + \cos(\tilde{\Omega}_{\text{toy}} t')\right)\left((\lambda y)^2 + \cos(\tilde{\Omega}_{\text{toy}} t'')\right)}{(1 + (\lambda y)^2)^2}.
    \end{aligned}
    $$
\end{itemize}
Thus, the correlator is
\begin{equation}
    C_{\lambda} (t', t'') = N \frac{(\lambda y)^2}{(1+(\lambda y)^2)^2} \left(1-\cos(\tilde{\Omega}_{\text{toy}} t')\right) \left(1-\cos(\tilde{\Omega}_{\text{toy}} t'')\right) + N\frac{1}{1+(\lambda y)^2} \sin(\tilde{\Omega}_{\text{toy}} t') \sin(\tilde{\Omega}_{\text{toy}} t'').\\
\end{equation}

This is a general expression for the correlator within the Born series approximation. The Taylor expansion of the correlator to second order in $\lambda$ yields
\begin{equation}
\begin{split}
    C_{\lambda} (t', t'') &= N \left( \sin(\tilde{\Omega}_{\text{toy}} t') \sin(\tilde{\Omega}_{\text{toy}} t'') \right. \\
    &+ \left. (\lambda y)^2 \left[ \left(1-\cos(\tilde{\Omega}_{\text{toy}} t')\right) \left(1-\cos(\tilde{\Omega}_{\text{toy}} t'')\right) - \sin(\tilde{\Omega}_{\text{toy}} t') \sin(\tilde{\Omega}_{\text{toy}} t'') \right] + \mathcal{O}(\lambda^4)\right) \,.
    \end{split}
\end{equation}
Inserting it in the Born expansion of the kernel in Eq. \eqref{eq:toykernel_bornexp}, we see that only the 0-th order term of the correlator contributes to 2nd order terms:
\begin{align}
    \kappa_{\text{toy}} (x,y) &= 1 - \lambda^2 (x-y)^2 \left(\frac{\Omega_{\text{toy}}}{2}\right)^2 N \int_0^T dt' \int_0^T dt'' \sin(\tilde{\Omega}_{\text{toy}} t') \sin(\tilde{\Omega}_{\text{toy}} t'') + \mathcal{O}(\lambda^4) \\ &= 1 - \tilde{\lambda}^2 (x-y)^2  N \sin^4 \left( \frac{\tilde{\Omega}_{\text{toy}} T}{2} \right) + \mathcal{O}(\lambda^4) \,, \label{eq:lastexpansion}
\end{align}
where $\tilde{\lambda}=2\delta /\tilde{\Omega}_{\text{toy}}$ and $\tilde{\Omega}_{\text{toy}} = \Omega_{\text{toy}} \sqrt{1+(\lambda y)^2}$ is the generalized Rabi frequency.\\

Notice that, in the perturbative regime where $\lambda\ll 1$, we can further consider the approximation $\tilde{\lambda} \approx \lambda$ and $\tilde{\Omega}_{\text{toy}} \approx \Omega_{\text{toy}}$ (all other deviations from this would lead to $\mathcal{O}(\lambda^4)$ contributions to Eq.~\eqref{eq:toykernel_bornexp}). This leads to a simplified expression of the toy kernel:  
\begin{equation}
    \boxed{
    \kappa_{\text{toy}} (x,y) = 1 - \lambda^2 (x-y)^2  N \sin^4\left( \frac{\Omega_{\text{toy}} T}{2} \right) + \mathcal{O}(\lambda^4).
    }
    \label{eq:simplifiedtoykernel}
\end{equation}
Interestingly, the result would remain exactly the same even if the perturbation were \emph{not} chosen to be a staggered field— that is, if we had taken $V = \sum_{j=1}^{N} \sigma_j^z$.
\end{widetext}

\subsection{Mean and variance of the single-feature toy kernel \label{supp-subseq:toymeanvar}}

The mean and variance of the kernel are determined by the statistical properties of the dataset $\mathcal{X}$, whose data points $x,y$ are i.i.d.~according to some probability distribution $P$. For the single-feature toy kernel $\kappa_{\text{toy}}(x,y)$ in Eq.\eqref{eq:simplifiedtoykernel}, we find the mean
\begin{align}
\begin{split}
    &\mathbb{E}_{x,y\in \mathcal{X}}[\kappa_{\text{toy}}(x,y)] =\\ &\;\;1 - N\lambda^2 \ \mathbb{E}_{x,y\in \mathcal{X}}\left[(x-y)^2\right] \sin^4\left( \frac{\Omega_{\text{toy}} T}{2} \right) \,, 
\end{split}
\end{align}
and variance,
\begin{widetext}
\begin{align}
    \text{Var}_{x,y\in \mathcal{X}}[\kappa_{\text{toy}}(x,y)] &= \mathbb{E}_{x,y\in \mathcal{X}}\left[\kappa_{\text{toy}}^2(x,y)\right] - \mathbb{E}_{x,y\in \mathcal{X}}\left[\kappa_{\text{toy}}(x,y)\right]^2 \\&= N^2\lambda^4 \ \left( \mathbb{E}_{x,y\in \mathcal{X}}\left[(x-y)^4\right] - \mathbb{E}_{x,y\in \mathcal{X}}\left[(x-y)^2\right]^2 \right) \sin^8\left( \frac{\Omega_{\text{toy}} T}{2} \right).
\end{align}
\end{widetext}
For a fixed time $T$ and given a probability distribution (with associated moments $\mathbb{E}_{x,y\in \mathcal{X}}\left[(x-y)^k\right]$, $k=1,2,\dots$), the mean and variance exhibit the following dependence in the system size $N$ and the encoding strength $\lambda$:
\begin{align}
    1 - \mathbb{E}_{x,y\in \mathcal{X}}[\kappa_{\text{toy}}(x,y)] &\propto N\lambda^2 \label{eq:meandep}\\
    \text{Var}_{x,y\in \mathcal{X}}[\kappa_{\text{toy}}(x,y)] &\propto N^2\lambda^4. \label{eq:vardep}
\end{align}
For $x,y$ following a uniform distribution over $[0,1]$, as in the main text, we have that $\mathbb{E}[x^k]=1/(k+1)$, meaning $\mu = \mathbb{E}[x] =1/2$ and $\sigma^2 = 1/12$. This leads to $\mathbb{E}_{x,y\in \mathcal{X}}\left[(x-y)^2\right] = 1/6$ and $\mathbb{E}_{x,y\in \mathcal{X}}\left[(x-y)^4\right] = 1/15$, giving prefactors of $1/6\approx 0.167$ and $7/180\approx 0.039$ in Eqs.~\eqref{eq:meandep} and~\eqref{eq:vardep}, respectively.

\section{Details on the classical simulation of \textsc{RydKernel} \label{supp-seq:classsim}}
\subsection{TEBD simulations of one-dimensional \textsc{RydKernel} \label{supp-subseq:tebdsim}}

The tensor network algorithm used to simulate the Rydberg-blockaded dynamics is based on a matrix-product-state (MPS) representation of the quantum state of the system. In this representation, the $N$-site quantum state vector is decomposed into a chain of $N$ tensors of rank 3, one for each site, such that each is expressed in the optimal basis regarding the bipartite entanglement content of the system. The side dimension of each tensor — called the bond dimension $\chi$ — which explicitly reflects this information, allowing for a compressed representation of the quantum state simply by truncating the tensor to a lower $\chi$ \cite{Schollwock2011}.

Trotterizing the unitary Hamiltonian evolution into a product of local unitaries allows for efficiently applying them on the MPS representation. This is the time-evolving block decimation (TEBD) algorithm. We use a second-order Trotterization of the NN Rydberg Hamiltonian, such that the evolution over a time interval $T$ is given by 
\begin{equation}
    U_{\text{Ryd}}(T) = \left(e^{-i\frac{dt}{2}H_A} e^{-idtH_B} e^{-i\frac{dt}{2}H_A} \right)^M + \mathcal{O}(Tdt^2\mathcal{C}) \,.   
\end{equation}
where $M=T/dt$ is the number of time steps, and $H_A$ ($H_B$) refer to the Hamiltonians acting on alternating even (odd) sites.

The trotterization error essentially depends on the product of $dt^2$ and the nested commutator of the noncommuting terms in the Hamiltonian: $\mathcal{C} \sim \left\lVert[H_B,[H_B,H_A]] \right\rVert \sim V_{i,i+1}^2$, the latter being of the order of the square of the blockade interaction \cite{Childs2021,Hahn2025}. Thus, the maximum time step is set by the inverse of the interaction energy scale, $dt \lesssim 1/V_{i,i+1}$. For the parameters in the paper that is, $\Omega=2$ and $V_{i,i+1} = 4.4 \Omega$, we use $dt=0.02$.

Another source of error in TEBD is the truncation of the bond dimension $\chi$ at each step. In our simulations, we use the more realistic NN Rydberg Hamiltonian rather than a low-energy approximation such as the PXP model. As a result, the dynamics can leak out of the constrained subspace defined by the Rydberg blockade. This leakage slightly increases the maximum bond dimension required to accurately simulate the dynamics. However, we find that a bond dimension $\chi=230$ for $N=21$ allows to keep the error in the fidelity kernel value $\leq 10^{-3}$ at $T=3.0 T_{\text{rev}}$. The computational work was performed using the Python Quimb package \cite{gray2018quimb}.

\subsection{Hardness of classical simulations \label{supp-subseq:hardness}}

\begin{figure*}[!tbp]
\includegraphics[width=\textwidth]{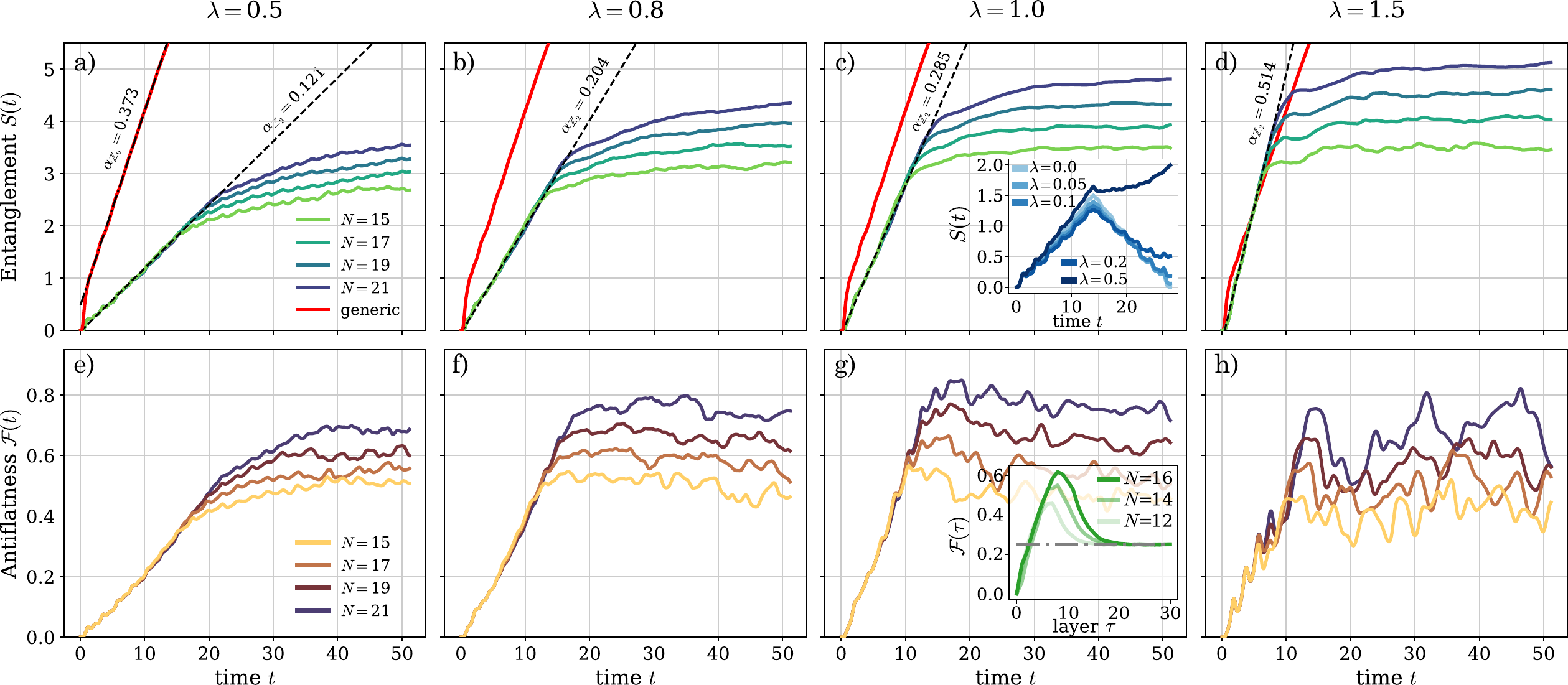} 
\caption{(a,b,c,d) Initial ballistic growth of entanglement entropy for the $\ket{\mathbb{Z}_2}$-Rydberg-blockaded quench dynamics for different values of the perturbation $\lambda$. Inset: Entanglement growth during \textsc{RydKernel} echo ($M=1$, $x-x'=1$) for different encoding strengths $\lambda$ in the perturbative regime, with $N=21$ and $T=3.0 T_{\text{rev}}$. (e,f,g,h) Time evolution of anti-flatness for different values of the perturbation $\lambda$. Inset: Antiflatness in random unitary circuit evolution (average over 100 instances), gray dashed-dotted line is $\mathcal{F}_{\text{Haar}}=0.25$. TEBD simulations with a maximum bond dimension of $\chi=3070$.}
\label{fig:smcompound4new}
\end{figure*}

\subsubsection{Entanglement growth}

Quantum states resulting from generic dynamics in one-dimensional quantum systems of size $N$ require $\chi = d^{N/2}$ exponential in the system size to be faithfully simulated ($d$ being the local effective dimension). This so-called volume-law entanglement, $S \propto N$, is reached after an initially ballistic entanglement growth regime, $S(t)= \alpha t$ with $\alpha$ system-size independent, of relatively short duration $t_{\text{bal}} \propto N$, after which entanglement entropy saturates. After $t_{\text{bal}}$, the bond dimension, and thus the memory required for the classical simulation, are exponential in the system size. The early-time regime, where MPS simulation is exact for arbitrarily large systems, ends as soon as $S(t)$ crosses the maximum simulatable entanglement $S_{\text{max}}$, which happens at $t_{\text{max}} = S_{\text{max}}/\alpha \sim \ln \chi$. In the Rydberg-blockaded chain this kind of generic dynamics arises from the $\ket{\mathbb{Z}_0}$ state for instance, and the entanglement entropy at saturation is roughly $S \sim N\ln \phi$, instead of $S \sim N\ln 2$, for this kind of constrained dynamics.

While the Rydberg-blockaded dynamics from the $\ket{\mathbb{Z}_2}$ state is not generic, it shows generic features as it leads to highly entangled states, as shown in the main text for zero detuning ($\lambda=0$). The main difference from generic dynamics was that saturation is not reached immediately, but only at later times. 

We provide additional insights on this behavior for stronger perturbations. In particular, in Figs. ~\ref{fig:smcompound4new}(a,b,c) we show that the initial ballistic growth of the entanglement entropy, $S(t)= \alpha_{\mathbb{Z}_2} t$, persists in the EC-free phase ($\lambda<1$) beyond the perturbative regime. We see that as $\lambda$ increases, so does the growth rate $\alpha_{\mathbb{Z}_2}$, until it reaches a value close to the generic one at the edge of the EC-free phase ($\lambda=1$). Consequently, volume-law entanglement is reached after a shorter time ($t_{\text{bal}}^{\mathbb{Z}_2}(\lambda>0.2) < t_{\text{bal}}^{\mathbb{Z}_2}(\lambda=0)$). We note also that for such high values of $\lambda$ the entanglement entropy saturates much faster at the theoretical maximum ($S_{\text{max}} = \frac{N}{2}\ln \phi$). As a result, the cost of classically simulating the $\ket{\mathbb{Z}_2}$ Rydberg-blockaded dynamics from the perspective of entanglement remains exponential in system size within the whole EC-free phase.

Beyond the EC-free phase (Fig. ~\ref{fig:smcompound4new}d), the ballistic growth rate increases further and, most importantly the entanglement saturates at a larger value than the one given by the Rydberg-blockade constraints, meaning that the dynamics has leaked outside this subspace. This reinforces the claim that the EC-free character of \textsc{RydKernel} is directly related to this constrained structure. 

Small encoding perturbation on the contrary has little impact on entanglement growth. It slightly reduces the entanglement entropy growth rate ($\lambda\lesssim0.2$), before increasing it again when stronger, as shown in the inset of Fig.~\ref{fig:smcompound4new}c. Most importantly, we see that, for $\lambda \gtrsim 0.1$, the disentanglement caused by the backward part of the Loschmidt echo dynamics is imperfect, potentially keeping entanglement high.


This analysis shows that the $\ket{\mathbb{Z}_2}$-Rydberg-blockaded dynamics leads to volume-law entanglement at relatively short times, rendering faithful classical MPS simulations out-of-reach for moderate sizes and times. Moreover, classically simulating the quantum kernel matrix for a dataset $\mathcal{X}$ amounts to computing $N_S (N_S+1)/2 \sim \mathcal{O}(N_S^2)$ matrix elements (since the kernel is symmetric) where $N_S = |\mathcal{X}|$ is the size of the dataset. This must be contrasted with the cost of computing the kernel on neutral atom quantum computers. In the absence of exponential concentration, estimating each matrix element simply amounts to measuring the occurrence of the $|\mathbb{Z}_{2} \rangle$ bitstring, which requires a number of shots polynomial in the system size to be faithfully evaluated. The rate of $\sim 1\text{ experiment/s}$ on current devices is limited by array loading and imaging, and can be enhanced by parallelizing on multiple atom-chains simultaneously, as well as fast imaging and continuous reloading.

\subsubsection{Nonlocal magic lower-bound}
\label{supp seq: NL magic}
In this section we provide (1) additional information on magic and its relationship with entanglement through anti-flatness of the entanglement spectrum, as well as (2) anti-flatness results for the perturbed Rydberg-blockaded dynamics and (3) a brief review of relevant results for the PXP dynamics already available in the literature~\cite{Smith2025}. 

While numerous measures of magic have been proposed~\cite{Bravyi-UQC-clifford-magic-PhysRevA.71.022316}, most remain computationally intractable even numerically. Stabilizer Rényi Entropies (SREs) stand out as efficiently computable magic monotones that admit closed-form analytical expressions in many settings~\cite{Leone-stabilizerentropy-monotone-PhysRevA.110.L040403}. Any operator on the $N$-qubit Hilbert space (of dimension $d = 2^N$) admits a decomposition over the $4^N$ dimensional Pauli operator string basis $\{P \in \mathbb{P}_N\}$, where each element takes the tensor product form 
$P=\bigotimes_{i=1}^{N} \sigma_i^{\alpha}$, with 
$\sigma_i^{\alpha}\in\{\mathds{1}, \sigma_i^{x}, \sigma_i^{y}, \sigma_i^{z}\}$ for $\alpha=0,1,2,3$ respectively. The operator of interest here is the density matrix of the evolving pure state, $\psi(t)= \dyad{\psi(t)}$, which decomposes as
\begin{equation}
    \psi(t) = \sum_{i=1}^{n_{\psi}} c_i(t)\, P_i,
\end{equation}
where the sum runs over the $n_\psi \leq 4^N$ Pauli strings 
and the coefficients $c_i(t)=\frac{1}{d}\operatorname{Tr}[\psi(t) P_i]$ are real, since both $\psi(t)$ and $P_i$ are Hermitian. The normalization $\sum_{i=1}^{n_{\psi}} |c_i(t)|^2 =1$ 
follows directly from the purity condition $\operatorname{Tr}[\psi^2]=1$, so that at any given time the $\{|c_i(t)|^2\}$ define a probability distribution over the Pauli strings composing the state. Stabilizer states are supported on a restricted subset of mutually commuting Pauli strings, all entering the decomposition with equal weight. Magic states, by contrast, spread across a much larger number of Pauli strings~---~potentially all $4^N$~---~with unequal weights, reflecting the multi-branching nature of the quantum dynamics required to generate them~\cite{Leone-SRE2022-PhysRevLett.128.050402, Turkeshi2025-nonstabilizerness-PhysRevB.111.054301}. The SRE is built on this observation: it quantifies the spread of the distribution $\{|c_i(t)|^2\}$ over Pauli strings, and thereby serves as a measure of magic.
The stabilizer 2-R\'{e}nyi entropy (SRE${_2}$) is most commonly used and for pure states it is defined as,
\begin{equation}
    \mathcal{M}_2(\ket{\psi}) = -\ln \sum_{P\in \mathbb{P}_N} \frac{1}{2^N}|\bra{\psi} P\ket{\psi}|^4 \; .  
\end{equation}
$\mathcal{M}_2(\ket{\psi}) = 0$ if and only if $\ket{\psi}$ is a stabilizer state, otherwise $\mathcal{M}_2(\ket{\psi}) > 0$. Moreover, it is invariant under Clifford operations. Finally, SRE${_2}$ is upper bounded by $N\log 2$ for $N$-qubit systems (and by $\log d$ in general, $d$ being the dimension of the full Hilbert space). Note that the related linear SRE, $\mathcal{M}_{\text{lin},2}(\ket{\psi})=1-e^{-\mathcal{M}_2(\ket{\psi})}$, can also be used if one seeks a strong monotone. By additivity of the SRE and monotonicity under Clifford operations, non-stabilizing power related to the $\mathcal{M}_2(\ket{\psi})$ provides a lower bound on the number of single-qubit $T$ gates required to prepare 
$|\psi\rangle$ from a stabilizer state, and equivalently on the number of single-qubit 
magic states that can be distilled from 
this state~\cite{Leone-SRE2022-PhysRevLett.128.050402,LiuWinter2022-manybodymagic-PRXQuantum.3.020333}.

It is important to note that magic alone does not determine classical simulation complexity. Hardness of simulation arises from the interplay between magic and entanglement: a highly magical but unentangled state remains efficiently simulable, while it is the conjunction of magic with Clifford operations that generates genuine classical hardness. For example, a product state of $N$ single-qubit magic states and an $N$-qubit Haar-random state carry a similar magic content, both scaling proportionally to $N$~\cite{Leone-SRE2022-PhysRevLett.128.050402}. Yet their simulation costs are vastly different: the product state is efficiently simulable due to its lack of entanglement, while the Haar-random state is exponentially hard to simulate classically, owing to the combination of maximal magic and maximal entanglement. This illustrates that magic is a necessary but not sufficient resource for classical hardness of simulation.

Non-local (NL) magic is precisely the component of magic that resides in the quantum correlations between different subsystems of a state, and as such it necessarily requires entanglement to exist~\cite{Zanardi2026-nonlocalSTAB-9x56-2b45, Cao2024_graviBR_magic_z3vr-w5c5}. Moreover, both quantities are intimately related to entanglement: NL 
magic is lower bounded by the non-flatness of the entanglement spectrum and upper bounded by the amount of entanglement in the system~\cite{Cao2024_graviBR_magic_z3vr-w5c5}. On the other hand, total magic in a system and entanglement are more broadly intertwined as complementary quantum resources~\cite{Tirrito_Antiflatness_PhysRevA.109.L040401,Frau2024-nonSTAB-entang-MPS-PhysRevB.110.045101, Dowling2026-EntMagicc7k1-xcwy}.

Considering an arbitrary bipartite pure state $\ket{\psi_{AB}}$, NL magic is thus defined as the minimal amount of magic contained in a bipartite state after optimally removing all ``local” magic in subsystems $A$ and $B$ separately, 
\begin{equation}
    \mathcal{M}_{2}^{\text{NL}} (\ket{\psi_{AB}}) = \min_{U_A \otimes U_B} \mathcal{M}_2 (U_A \otimes U_B \ket{\psi_{AB}}) \ ,
\end{equation}
where minimization is performed over all possible local unitary transformations on each subsystem. 

A remarkable result in the resource theory of magic  establishes that the magic of a pure state is proportional to the average deviation of its entanglement spectrum from flatness — the so-called \emph{anti-flatness} — taken over its full Clifford orbit, and this holds independently of the choice of bipartition~\cite{Tirrito_Antiflatness_PhysRevA.109.L040401, Cao2024_graviBR_magic_z3vr-w5c5}. Equivalently, a state possesses magic if and only if its entanglement spectrum is not flat~\cite{Tirrito_Antiflatness_PhysRevA.109.L040401}. Pure stabilizer states, despite being capable of carrying substantial entanglement, necessarily possess a flat entanglement spectrum, and are therefore magic-free.

NL magic has a particularly tight relationship to entanglement: $\mathcal{M}_{2}^{\text{NL}} (\ket{\psi_{AB}}) = 0$ iff the entanglement spectrum is flat, which is the case for magic product states and stabilizer states (Lemma 1 in \cite{Cao2024_graviBR_magic_z3vr-w5c5}), and is positive otherwise. Non-local magic therefore precisely captures the non-stabilizer content of the correlations between subsystems. Consequently, it can be probed through the so-called anti-flatness of the entanglement spectrum, which provides a rigorous lower bound on NL magic (Theorems 2 and 3 in \cite{Cao2024_graviBR_magic_z3vr-w5c5}).

Anti-flatness is measured by computing the variance of the reduced density matrix over itself,
\begin{equation}
    \mathcal{F}(\rho_A) = \text{Tr}(\rho_A^3) - \text{Tr}^2(\rho_A^2) \ .   
\end{equation}
Dividing this expression by the mean squared, $\text{Tr}^2(\rho_A^2)$, leads to the rescaled anti-flatness used in the main text. This yields a coefficient of variation — a standardized measure of dispersion of a probability distribution that normalizes by the mean of the distribution — allowing meaningful comparison between entanglement spectra whose means differ widely. This is precisely the situation arising in quantum many-body dynamics, where the support of the entanglement spectrum grows rapidly in time while the mean (purity) simultaneously decreases.

In \cite{Smith2025}, the magic ( $\text{SRE}_2$ ) generated along the effective PXP dynamics from both the $\ket{\mathds{Z}_{2}}$ and the thermalizing $\ket{0}$ state was numerically evaluated using TDVP with replica-MPS with $\chi=16$ for $N=51$ qubits and up to $t\approx 1T_{\text{rev}}$. For $\ket{0}$ the magic shows a short rapid increase at very early times, followed by a slower but steady increase with little fluctuations, while for $\ket{\mathbb{Z}_2}$ the magic grows more slowly, exhibiting repeated three-peak oscillatory pattern (small-large-small) over each half-revival. Moreover, using analytics in a projected $\chi=2$ MPS manifold, they show that some amount of long-range SRE (a quantity closely related to non-local magic) is developed along both evolutions and that it is tied to beyond-nearest-neighbor correlations. In particular, for $\ket{\mathbb{Z}_2}$, long-range SRE exhibits periodic peaks aligned with the central (large) peak of the global SRE three-peak pattern.  

However, comparison with the TDVP numerics for the global magic shows that this analytical technique only captures the periodic features of the SRE evolution (global and long-range) but not the overall growth in time of this quantity. Thus, all the  SRE results go back to zero periodically in this framework as it provides only a semiclassical representation of the state’s non-local magic. Additionally, eventhough replica MPS method  is among the most efficient for evaluating SRE, it scales as $\chi^{12}$, making later times in the evolution extremely hard to reach.

In Figs. ~\ref{fig:smcompound4new}(e,f,g) we provide additional insights on the behavior of non-local magic probed by anti-flatness in the $\ket{\mathbb{Z}_2}$ Rydberg-blockaded dynamics for strong perturbations. In particular, anti-flatness initially increases at a rate independent of system size, before entering an intermediate regime in which it becomes proportional to 
system size, $\mathcal{F} \sim N$. This scaling persists beyond the perturbative regime throughout the entire EC-free phase ($\lambda < 1$). Since anti-flatness 
provides a lower bound on NL magic, the classical complexity of simulating the dynamics is consequently expected to remain exponential in system size within the 
EC-free phase.

Additionally, the initial growth rate increases with $\lambda$. We notice, however, that for $\lambda \to 1$, the anti-flatness reaches its maximum within the investigated time window, before visibly decreasing while exhibiting larger fluctuations. For $\lambda>1$ beyond the EC-free phase (Fig. ~\ref{fig:smcompound4new}h), the initial growth phase clearly shrinks, and while a system-size-dependent peak of smaller magnitude is visible at intermediate times, it is directly followed by large fluctuations of the anti-flatness.

For reference, we show the anti-flatness evolution in random unitary circuits (Inset of Fig. ~\ref{fig:smcompound4new}g). A system-size-dependent peak is clearly visible following the initial system-size-independent growth, after which the anti-flatness converges to a lower but non-zero universal value, $\mathcal{F}_{\text{Haar}} = 0.25$, at times $O(\log(N))$~\cite{TurkeshiRQC2025-Natcomm}. 

It is worth noting that the real amount of non-local magic generated at times where the entanglement spectrum is far from flat is expected to be much larger than the anti-flatness (possibly exponentially in the system size). This is what we expect for most investigated time of the Rydberg-blockaded dynamics. However, in the near-flat limit, concretely near $t=0$ (if the initial state is a computational product state) or when the system has fully thermalized and subsystems are almost maximally mixed, then the anti-flatness is expected to be proportional to non-local magic, $\mathcal{M}_2^{\text{NL}} (\ket{\psi_{AB}}) \approx \frac{\mathcal{F}(\rho_A)}{\text{Tr}^2(\rho_A^2)}$~\cite{Cao2024_graviBR_magic_z3vr-w5c5}.

\section{Details on the experimental implementation of \textsc{RydKernel} on analog neutral-atom simulator \label{supp-seq:implementation}} 

\begin{figure*}[b]
\includegraphics[width=0.8\textwidth]{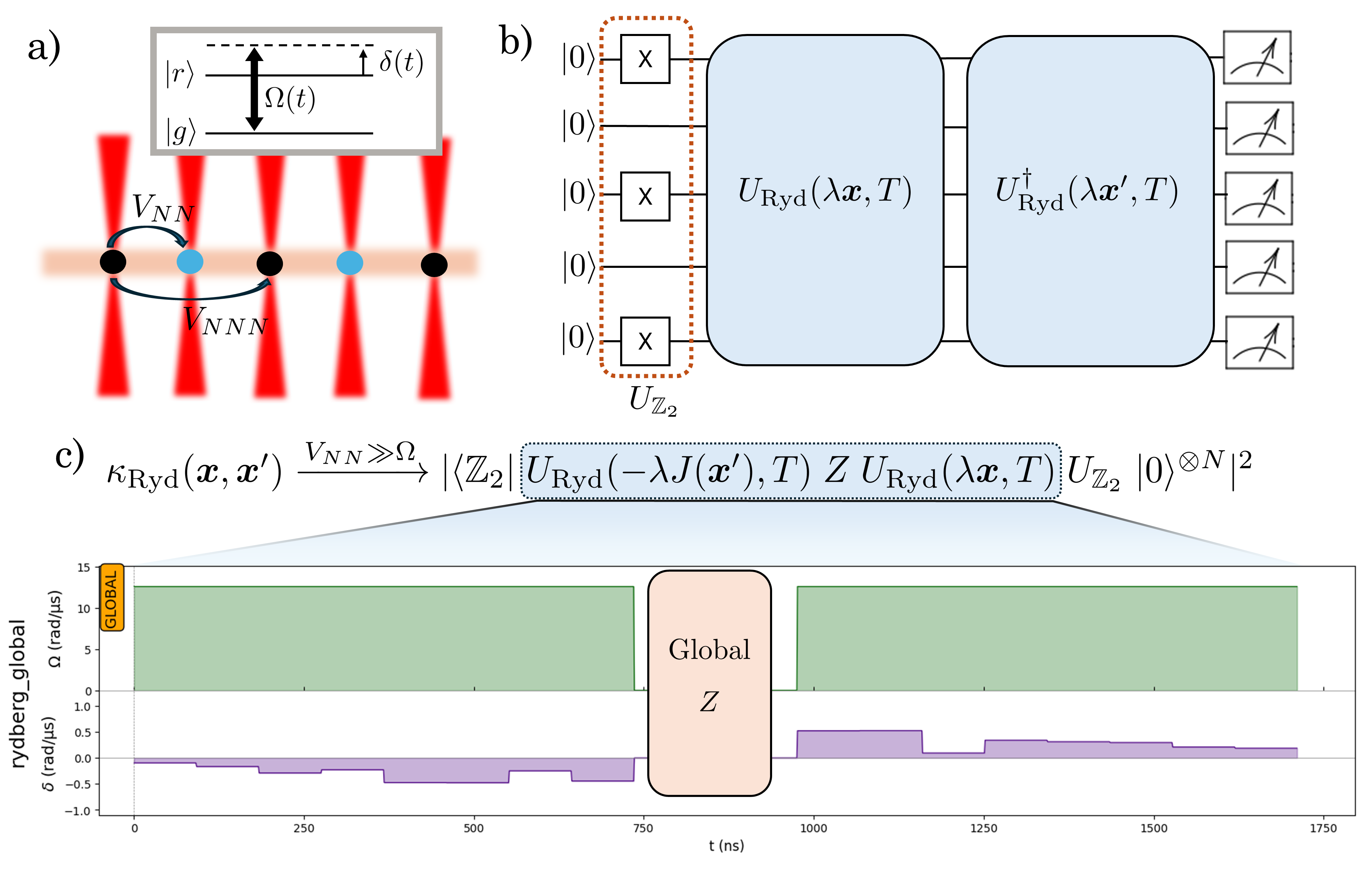} 
\caption{(a) Rydberg tweezer experimental platform where our proposal can be enacted, consisting of trapping lasers (beige) and lasers that are nearly resonant to the $\ket{g} \leftrightarrow \ket{r}$ transition between ground and Rydberg excited levels (red), with Rabi frequency $\Omega(t)$. The laser can be detuned away from resonance, such that $\omega_{\text{laser}}(t) = \omega_{gr} + \delta(t)$. The trapped atoms interact through strong Rydberg-blockaded interactions. (b) Visual representation of \textsc{RydKernel} as a digital-analog quantum circuit. The $U_{\mathds{Z}_2}$ blocks represent a digital preparation of the $|\mathds{Z}_2 \rangle$ state through local gates. The blue blocks represent the forward and backward evolution  encoding data points and realizing the Loschmidt echo. (c) Visualization of the laser pulse sequence implementing the Loschmidt echo protocol for $T=1.0 T_{\text{rev}}$ and $M=8$-feature data points. The approximate time-reversal is realized through a global $Z$ gate followed by a second forward evolution $U_{\text{Ryd}}(-\lambda J(\boldsymbol{x}'), T)$, as explained in the text. Pulse sequences are drawn using Pulser, a Python package for designing and simulating pulse sequences on programmable neutral-atom arrays\cite{Silverio2022}.}
\label{fig:rydbergsetup}
\end{figure*}

\subsection{Analog neutral-atom quantum computers (NAQC) \label{supp-subseq:naqc}}

NAQCs based on Rydberg-atom arrays are a leading candidate for analog quantum computation due to their remarkable performance regarding programmable connectivity, high-fidelity quantum operations and readout, and decoherence time.

These machines operate by trapping alkali atoms in a vacuum chamber using an optical tweezer array. Spatial light modulators allow for positioning atoms in arbitrary spatial configurations with a typical interatomic distance of $a\sim 5\mu m$. Of particular interest for our proposal are 1D chains and 2D grids, which have been demonstrated with systems of up to $N=6000$ atoms~\cite{Manetsch2025}. A technologically mature choice is rubidium ${}^{87}$Rb atoms for which the qubit ground and excited states can be encoded in the atomic ground and highly excited Rydberg states, respectively, as $|g\rangle \equiv |5S_{1/2}, F=2, m_F = 2\rangle$ and $|r\rangle\equiv |60S_{1/2}, m_J = 1/2\rangle$. The transition $|g\rangle \leftrightarrow |r\rangle$ is driven via a tunable laser drive, leading to the local Hamiltonian ($\hbar = 1$)
\begin{equation}
    H_{\text{loc}}(t) =\sum_{i = 1}^N \left( \frac{\Omega(t)}{2} \sigma_i^x - \delta(t) \hat{n}_i \right) \;,
    \label{eq:locrydham}
\end{equation}
where $\Omega(t)$ is the Rabi frequency, $\delta(t)$ is the detuning resonance with respect to the transition, $\sigma_i^x = |r\rangle_i\langle g|_i + |g\rangle_i\langle r|_i$ and $\hat{n}_i =(1 + \sigma_i^z)/2 = |r\rangle_i\langle r|_i$.

A crucial feature of this kind of system is that two atoms in the Rydberg state at sites $i$ and $j$ can be easily brought to a regime of strong interactions, as described by the interaction Hamiltonian  
\begin{equation}
    H_{\text{int}} = \sum_{i < j} \frac{C_6}{r_{ij}^6} \hat{n}_i \hat{n}_j \; ,
    \label{eq:intrydham}
\end{equation}
where the interaction strength $C_6$ is a constant set by the chosen Rydberg level and $r_{ij}$ is the real-space distance between atoms. When the interaction is much larger than the magnitude of the Rabi drive, $\Omega \ll C_6 / r_{ij}^6$, pairs of atoms that are closer than the ``blockade radius”, $r_{ij} \ll R_b$ ( $R_b = (C_6/\Omega)^{1/6}$) are energetically prohibited from simultaneously occupying their Rydberg state. This so-called Rydberg blockade regime allows for the generation of highly entangled states~\cite{Shaw2024} and the weak-ergodicity breaking phenomenon~\cite{bernien2017probing,bluvstein2021controlling} used in \textsc{RydKernel}. The Hamiltonians in Eqs.~\eqref{eq:locrydham} and~\eqref{eq:intrydham} together form the Rydberg-atom Hamiltonian (Eq.(3) in the main text).

As an illustration, let us consider an experimental setup consisting of a 1D chain with atom spacing $a = 5 \mu m$ and the Rydberg level $n_R =60$. Then, $C_6/\hbar \simeq 2\pi \times 137 \text{GHz}\cdot \mu \text{m}^6$ and $r_{ij} = a |i-j|$, leading to interaction strengths $V_{i,i+1} \simeq 8.82 \text{MHz}$ and  $V_{i,i+2} \simeq 0.138 \text{MHz}$ for the nearest- and next-nearest neighbor (NNN) terms, respectively. A visual representation of this experimental arrangement can be seen in Fig.~\ref{fig:rydbergsetup}(a). In commercial devices, typical maximum Rabi amplitudes $\Omega_{\text{max}}$ are on the order of $2\ \mathrm{MHz}$~\cite{quera_device, PasqalOrion}, while academic setups have achieved values as high as $7\ \mathrm{MHz}$~\cite{shaw2024benchmarking}. The detuning can easily reach $7\ \mathrm{MHz}$. Both $\Omega$ and $\delta$ typically exhibit errors between $1\%$ and $5\%$, depending on the calibration of the experimental apparatus. 

Finally, these Rydberg-atom platforms are subject to several sources of decoherence, in particular local relaxation and dephasing characterized by two decoherence times $T_1$ and $T_2$. For the $n_R=60$ Rydberg level, one can typically have $T_1 \sim 100 \mu s$ and $T_2 \sim 4.5 \mu s$.

\subsection{\textsc{RydKernel} as a Loschmidt echo or SWAP test \label{supp-subseq:leprotocol}}

\begin{figure*}
    \centering
    \includegraphics[width=0.8\linewidth]{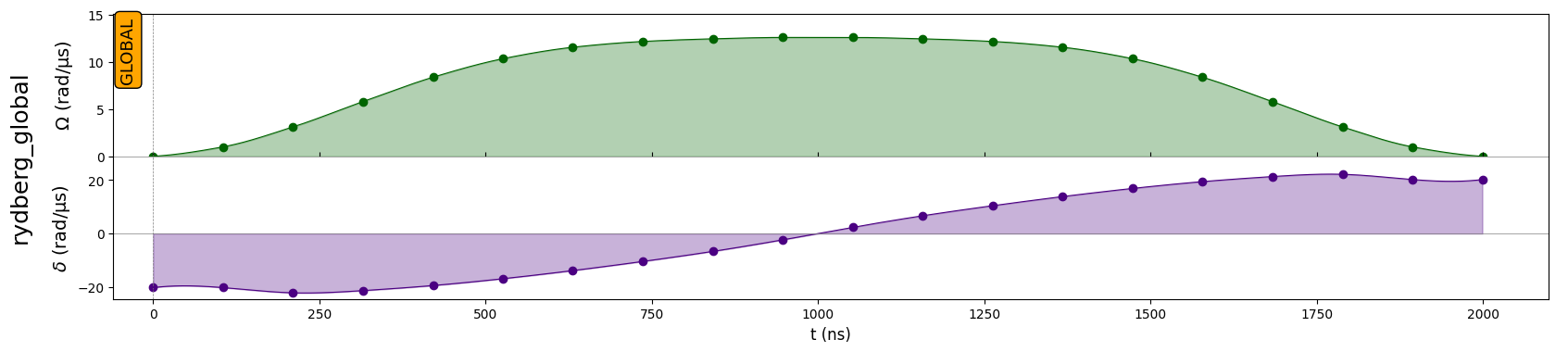}
    \caption{Annealing pulse sequence for the approximate preparation of the $|\mathds{Z}_{2}\rangle$ state in $T_{\text{prep}}= 2\mu \text{s}$.}
    \label{fig:supp-adiab-pulse-Z2}
\end{figure*}

A concrete realization of \textsc{RydKernel} can be implemented using a Loschmidt echo protocol, as described in the main text. The procedure consists of: \textbf{(1)} preparing the $\lvert Z_2 \rangle$ state, \textbf{(2)} performing a forward evolution that embeds the first data point, \textbf{(3)} applying an approximate time-reversed evolution that embeds the second data point, and finally \textbf{(4)} measuring in the computational basis. This implementation can be expressed as
\begin{align}
    \kappa_{\text{Ryd}}&(\boldsymbol{x}, \boldsymbol{x}') =\\ \Big|\underbrace{\langle \mathbb{Z}_2 |}_{\boldsymbol{(4)}} \; &\underbrace{U_{\text{Ryd}}(-\lambda J(\boldsymbol{x'}), T)\;Z}_{\boldsymbol{(3)}} \; \underbrace{U_{\text{Ryd}}(\lambda \boldsymbol{x}, T)}_{\boldsymbol{(2)}} \; \underbrace{U_{\mathbb{Z}_2}}_{\boldsymbol{(1)}} \; |0\rangle^{\otimes N} \Big|^2 \nonumber
\end{align}
and the details of each step is provided below.\\

\textbf{(1) $|\mathbb{Z}_{2}\rangle$-state preparation.}\\ 

\noindent After filling the array of traps with atoms in their ground state, different methods can be used to prepare the $|\mathbb{Z}_2\rangle = U_{\mathbb{Z}_2} |0\rangle^{\otimes N}= |rgrg\cdots\rangle$ initial state. The first is annealing which relies purely on global control of the lasers. Preparation times of $\sim 2000\text{ns}$ have been demonstrated on 1D chains \cite{Bernien2017} and a possible annealing schedule on $\Omega(t)$ and $\delta(t)$ is shown in Fig.~\ref{fig:supp-adiab-pulse-Z2}. The second method is to use semi-local addressing of the laser's detuning, forcing half of the atoms to stay in $| g \rangle$ (through the use of a Detuning Map Modulator \cite{goswami2024solving, de2025demonstration}), while each of their nearest neighbor is flipped to $| r \rangle$ via the application of a global $\pi$-pulse on $\Omega(t)$. Applying this  results in an effective staggered $X$ gate (similar to Fig.~\ref{fig:rydbergsetup}b) and can be implemented in $\sim 500\text{ns}$ ~\cite{Bluvstein2022}. In Ref.~\cite{Xiang2024} this protocol was used and a preparation fidelity of $\sim 49 \%$ was attained for a $N=25$ atom chain. Using feedback loops to optimize pulses may improve this fidelity ~\cite{chevallier2024variational}. Finally, note that while the use of local digital $X$ gates is possible in principle (Fig.~\ref{fig:rydbergsetup}b), this kind of digital-analog technology is not available on current commercial hardware.\\

\textbf{(2) Forward evolution encoding data point $\boldsymbol{x}$.}\\ 

\noindent $U_{\text{Ryd}}(\lambda \boldsymbol{x}, T)$ is implemented simply by placing the atoms in the Rydberg-blockade interacting regime for a time $T$, with the detuning perturbation parameterized by the renormalized data, $\delta = \Omega\lambda x/2$, as explained in the main text. While implementing an $M$-feature data vector $\boldsymbol{x}$ can theoretically be thought of as $M$ successive unitaries, $U_{\text{Ryd}}(\lambda\boldsymbol{x}, T) = \prod_{m=1}^{M} U_{\text{Ryd}}(\lambda x_m, T/M)$, in practice a continuous laser pulse is applied corresponding to a single unitary with time-dependent detuning. The latter takes the form of a step-function parameterized by a component $x_m$ of $\boldsymbol{x}$, each step of duration $T/M$. Fig.~\ref{fig:rydbergsetup}c shows an example of such a pulse sequence for a random $M=8$ data vector $\boldsymbol{x}$.\\

\textbf{(3) Approximate time-reversal encoding data point $\boldsymbol{x'}$.} \\

\noindent The backward evolution $U_{\text{Ryd}}^{\dagger}(\lambda \boldsymbol{x}', T)$ is not natively implemented on Rydberg-atom platforms. To achieve it, we can exploit the fact that the particle-hole symmetry operator $Z = \left(\prod_{i=1}^{N} \sigma_i^{(z)}\right)$ anticommutes with the low-energy effective description of the nearest-neighbor Rydberg hamiltonian in the Rydberg blockade regime, the PXP model \cite{Turner_2018} (see Sec.~\ref{toymodephysics}), to implement an approximate time-reversal evolution. This translates into a sign flip of the effective Hamiltonian, $Z H_{PXP} Z = - H_{PXP}$, while the effect is trivial on the detuning perturbation since $[Z,\hat{n}_i] = 0$. Thus, this allows for redefining the backward echo as an approximate echo, $U_{\text{Ryd}}^{\dagger}(\lambda \boldsymbol{x}', T) \approx  Z U_{\text{Ryd}}(-\lambda J(\boldsymbol{x}'), T) Z$, where $J(x'_1,..., x'_M) \rightarrow x'_M, ..., x'_1$ reverses the order in which data-vector components are encoded. The global $Z$ gate is implemented as a global $\pi$ phase shift in the neutral-atom setting, and the last $Z$ can be omitted since measurements are done in the computational basis. Such a protocol is described in ~\cite{Xiang2024}.

The global $Z$-gate does not reverse the sign of the next-nearest-neighbor interaction terms in the Rydberg Hamiltonian.
Even though the magnitude of these terms is small in the Rydberg-blockade regime ($V_{i, i+1} \gg \Omega \gg V_{i, i+2}$), their magnitude is a fixed fraction of the nearest-neighbor interaction---$V_{i, i+2} = V_{i, i+1}/2^6 = V_{i, i+1}/64$ for the 1D chain (Eq.\eqref{eq:intrydham})---leading to potentially detrimental effects on kernel properties after an encoding time scale $\sim 1/V_{i, i+2}$.\\

\textbf{(4) Measurement in the computational basis.} \\ 

\noindent Measurement in the computational basis is performed via fluorescence imaging, yielding a bitstring. A polynomial number of shots is sufficient to obtain an accurate estimate of the occurrence of the $\lvert \mathbb{Z}_2 \rangle$ state.

\subsection{SWAP-test protocol}
\begin{figure}[!tbp]
\includegraphics[width=0.95\textwidth]{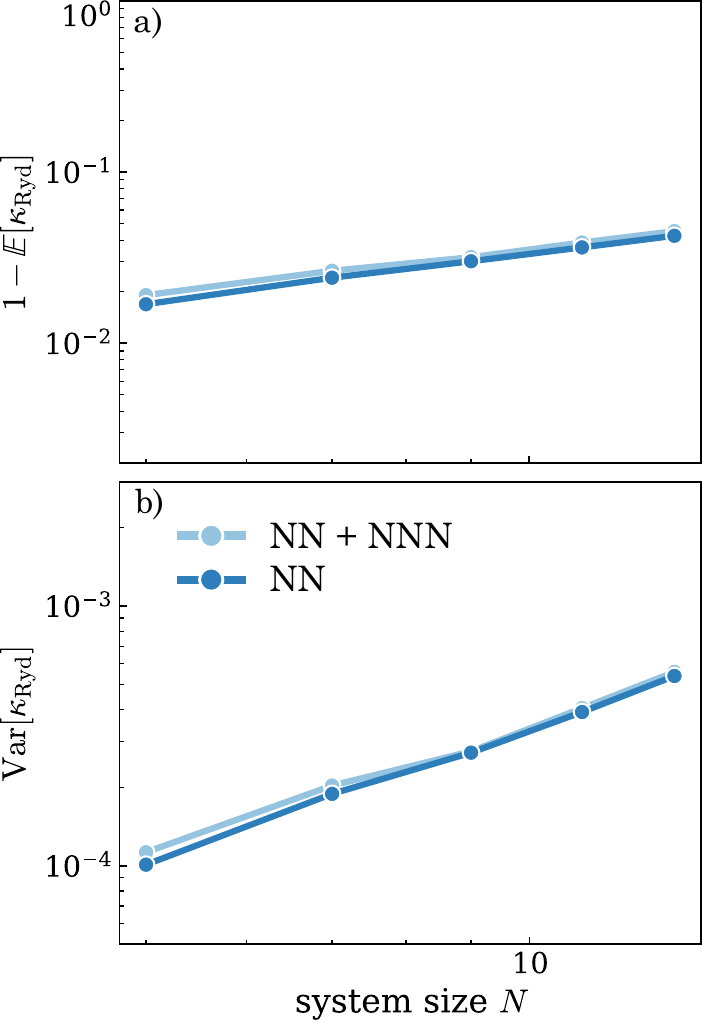} 
\caption{Effect of the next-nearest-neighbor interactions on \textsc{RydKernel}'s moments ($\lambda=0.2$, $N\in [5,13]$, $T=2.0 T_{\text{rev}}$ and $M=8$ features random dataset of size $|\mathcal{X}|=10$). (a) Mean of \textsc{RydKernel}. (b) Variance of \textsc{RydKernel}. Simulations are exact.}
\label{fig:smcompound5}
\end{figure}

The issue of implementing a backwards time evolution can be circumvented by implementing \textsc{RydKernel} using a SWAP-test protocol on a doubled system, rather than via the Loschmidt echo. This SWAP test has already been demonstrated experimentally in~\cite{Bluvstein2022}, although it requires transferring the $\ket{r}$ state to a hyperfine state $\ket{h}$, where digital operations can be performed, thereby introducing an additional source of infidelity. 
We assess the effect of the NNN terms on this implementation of \textsc{RydKernel} by performing exact simulations of the mean and variance on a random dataset. These results are shown in Fig.~\ref{fig:smcompound5}. We see that the inclusion of NNN terms does not affect the performance of the kernel and that exponential concentration remains absent.\\

\subsection{Concrete timescales for a minimal implementation of \textsc{RydKernel} \label{supp-subseq:concretetimescales}}

We now provide a quantitative estimate for a minimal experimental realization of \textsc{RydKernel}. We assume a Rydberg-atom quantum computer with semi-local control. A chain of atoms is prepared with atom-spacing  $a = 5 \mu$m, van der Waals interaction strength $C_6/\hbar = 2\pi \times 137 \text{GHz}\cdot \mu \text{m}^6$ and Rabi frequency $\Omega_{\text{max}} = 2\pi\times 2 \text{MHz}$. With these parameters, the revival time is $T_{\text{rev}} = 740 \text{ns}$ (independent of system size).

For the $|\mathbb{Z}_2\rangle$ state initialization we assume a preparation time of $T_{\text{prep}} = 500 \text{ns}$ based on the shorter semi-local method. Then, considering an encoding time of 2 revivals, $T = 2T_{\text{rev}} = 1480 \text{ns}$, for each of the unitaries embedding data points $x$ and $x'$, and a duration of $T_{Z} = 200 \text{ns}$ for the global $Z$ gate squeezed in between them, we end up with a total duration of $T_{\text{total}} = 3660 \text{ns}$ for the whole approximate Loschmidt echo protocol. This comfortably fits into the expected coherence time of $T_2 \simeq 4500 \text{ns}$ observed for this kind of quantum device. We note, however, that other setups, such as the one presented in Ref.~\cite{shaw2024benchmarking}, offer shorter $a$ and higher $\Omega_{\text{max}}$, and are therefore able to reach total evolution times of $T \simeq 10.4 T_{\text{rev}}$ with 60 atoms, which would correspond to encoding times $T \sim 4-5 T_{\text{rev}}$.

\section{Additional classification results on the IRIS dataset \label{supp-seq:irisclassif}}

\begin{figure}[!tbp]
\includegraphics[width=\textwidth]{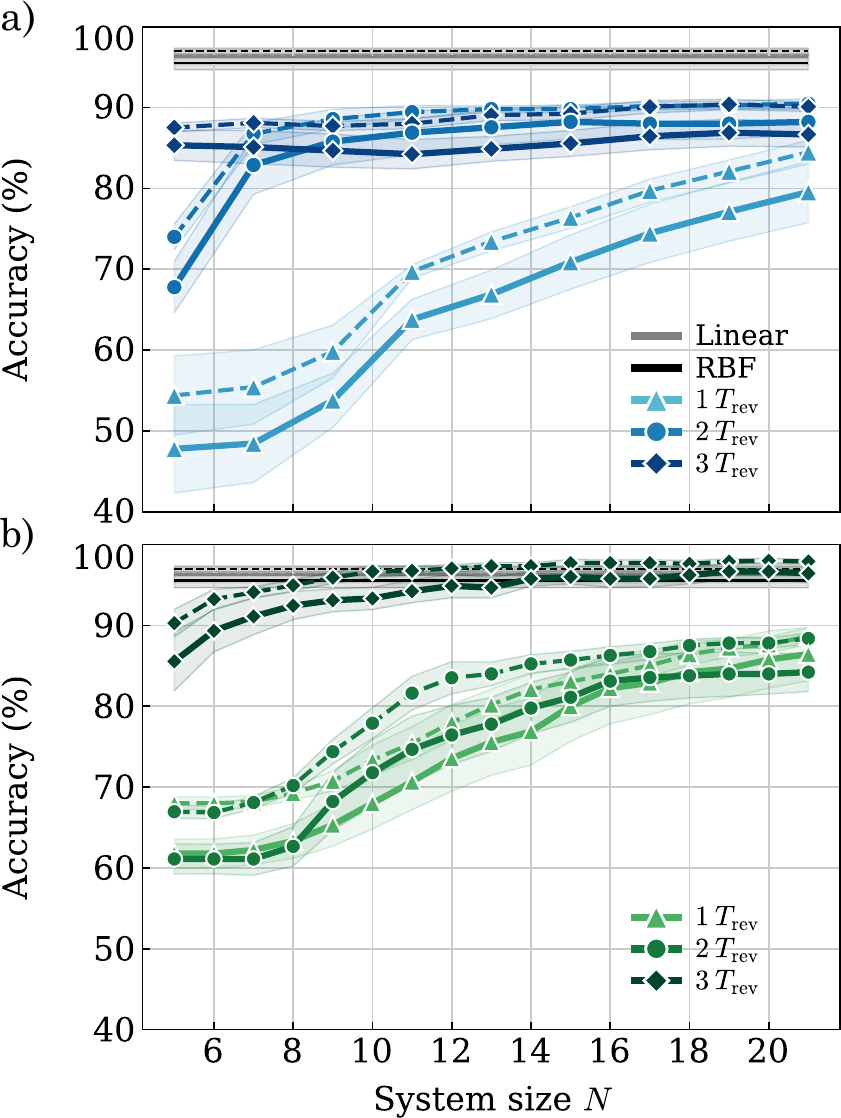} 
\caption{Classification performance of \textsc{RydKernel} (a) and the toy kernel (b) on the IRIS dataset at different encoding times $T=nT_{\text{rev}}$ ($n=1,2,3$) compared against the classical Linear and RBF kernels. Solid (dashed) lines correspond to test (resp. train) accuracies. We use $70\%$ of the dataset for training and $30\%$ for testing. Encoding strength of both $\kappa_{\text{Ryd}}$ and $\kappa_{\text{toy}}$ is set to $\lambda = 0.2$. Shaded regions represent the standard deviation resulting from 10-fold cross-validation.}
\label{fig:rydk_toyk_fixlambda}
\end{figure}

Figures~\ref{fig:rydk_toyk_fixlambda}(a,b) show the classification accuracy of \textsc{RydKernel} and the SU(2) toy kernel on the IRIS dataset as a function of system size with $\lambda$ fixed at 0.2. At $T = T_{\text{rev}}$, both kernels behave similarly, with accuracy increasing steadily with $N$.
At longer encoding times, such as $T = 3T_{\text{rev}}$, both kernels achieve high classification accuracy with little dependence on system size — though the toy kernel reaches a notably higher ceiling ($\sim 96\%$) compared to \textsc{RydKernel} ($\sim 87\%$). The picture changes substantially at $T = 2T_{\text{rev}}$, where the two kernels exhibit qualitatively distinct behavior.

\begin{figure}[!tbp]
\includegraphics[width=\textwidth]{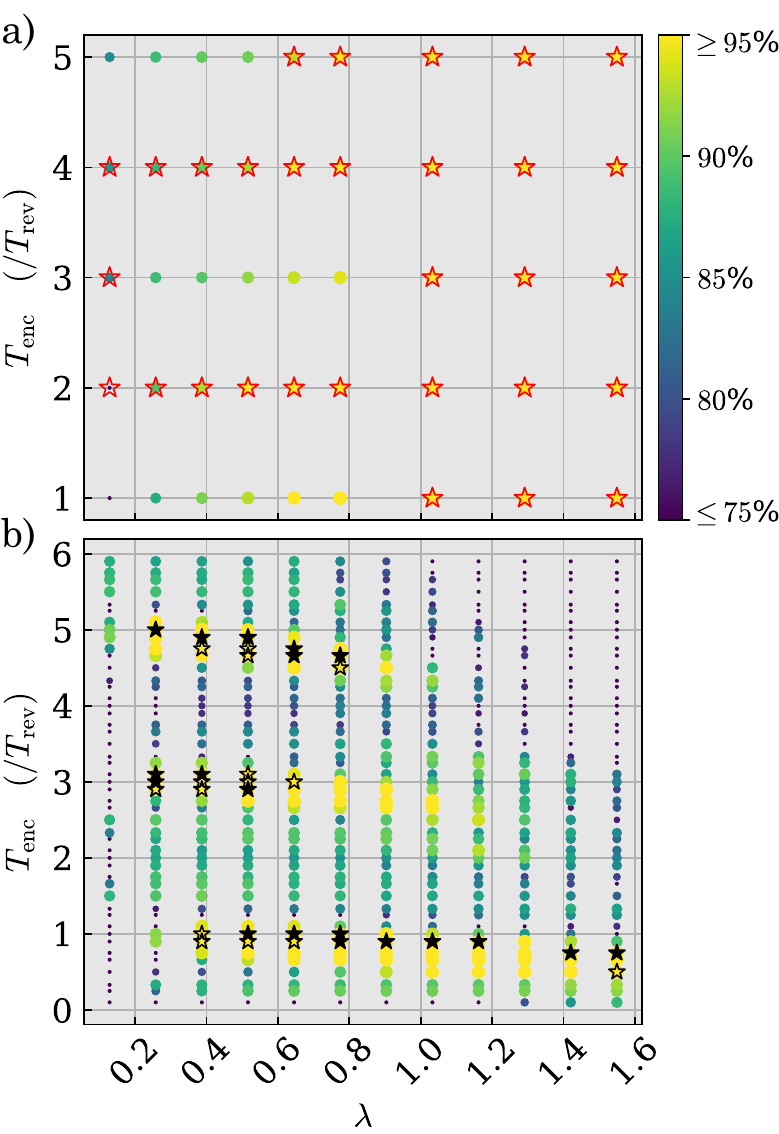} 
\caption{Classification accuracy at $N=15$ qubits for both \textsc{RydKernel} (a) and the toy kernel (b) on the IRIS dataset for a range of different encoding strengths $\lambda$ and encoding times $T$. The color and size of each dot both correspond to the test accuracy value. Accuracies lower or equal than $75\%$ (conventional threshold for successful classification) are not graphically differentiated, as well as accuracies above $95.0\%$ (accuracy of the lowest performing RBF classical kernel). (a) Red stars are the points for which \textsc{RydKernel} accuracy is better than toy kernel's. (b) Empty black stars are the points for which toy kernel's accuracy is better than $95\%$, filled black stars are the points for which toy kernel's accuracy is better than $96.6\%$ (best performing Linear classical kernel). $\lambda$ values are chosen such that $\lambda\propto 1/\sqrt{N}$. We use $70\%$ of the dataset for training and $30\%$ for testing.}
\label{fig:rydk_toyk_phasediag15}
\end{figure}

\begin{figure}[!tbp]
\includegraphics[width=\textwidth]{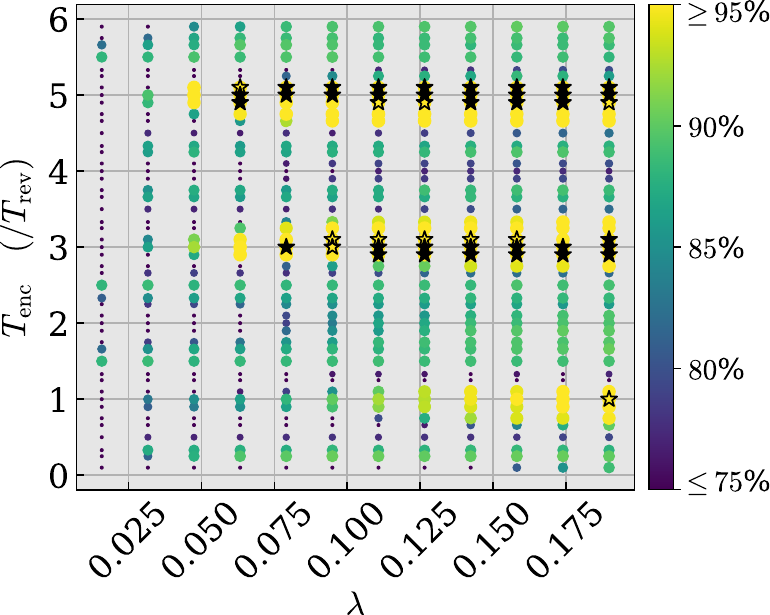} 
\caption{Classification accuracy at $N=1000$ qubits for the toy kernel on the IRIS dataset for a range of different encoding strengths $\lambda$ and encoding times $T$. Same legend as in Fig.\ref{fig:rydk_toyk_phasediag15}.}
\label{fig:toyk_phasediag1000}
\end{figure}

The degraded performance of the toy kernel at $T = 2T_{\text{rev}}$ is not meaningfully recovered by rescaling the detuning $\lambda \propto 1/\sqrt{N}$, as shown in the main text. This behavior appears to generalize: Figs.~\ref{fig:rydk_toyk_phasediag15} and~\ref{fig:toyk_phasediag1000} suggest that poor classification performance persists at other multiples of $T_{\text{rev}}$.

Figures~\ref{fig:rydk_toyk_phasediag15}(a,b) show the classification accuracy of \textsc{RydKernel} and the SU(2) toy kernel over a range of $(\lambda, T)$ at fixed $N = 15$. Since $\lambda > 1  (\geq\lambda_c)$ corresponds to the exponential concentration (EC) regime for \textsc{RydKernel}, we restrict our analysis to $\lambda < 1$. On the Iris dataset, \textsc{RydKernel} outperforms the toy kernel near even multiples of $T_{\text{rev}}$, while the toy kernel recovers strong performance near odd multiples of $T_{\text{rev}}$, occasionally matching or exceeding the best classical baselines (linear kernel). In Fig.~\ref{fig:rydk_toyk_phasediag15}(a), yellow stars with red borders mark parameter values where \textsc{RydKernel} outperforms the toy kernel. In Fig.~\ref{fig:rydk_toyk_phasediag15}(b), empty black stars indicate points where the toy kernel exceeds $95.0\%$ accuracy, while filled black stars mark points where it surpasses $96.6\%$ — the accuracy of the best classical kernel. In both panels, $\lambda$ values are chosen with $\lambda \propto 1/\sqrt{N}$. 

Figure~\ref{fig:toyk_phasediag1000} extends the phase diagram of the toy kernel to $N = 1000$ qubits on the Iris dataset, spanning a range of different encoding strengths $\lambda$ and encoding times $T$ with the legends being the same as in Fig.~\ref{fig:rydk_toyk_phasediag15}. 
The accuracy landscape differs slightly from the $N = 15$ case, owing to the different range of encoding strengths $\lambda$ sampled. At short encoding times, high accuracy ($> 95\%$) is only achieved at larger $\lambda = 0.175$, whereas near odd multiples of $T_{\text{rev}}$ such as $3T_{\text{rev}}$ and $5T_{\text{rev}}$, strong performance persists even at small and modest encoding strengths. Nevertheless, no meaningful improvement in classification accuracy is gained by increasing $N$, consistent with the classical simulability of the toy kernel.

All reported classification accuracies for both \textsc{RydKernel} and the toy kernel are obtained via 10-fold cross-validation.

\section{Details on the classical machine learning methods \label{supp-seq:mltasks}}

In this section, we detail the classical methods we use to (a) deploy a given kernel $\kappa(\boldsymbol{x}, \boldsymbol{x'})$ in a support vector machine (SVM) for data classification (see Sec.~\ref{supp-subseq:svm}) and (b) compare \textsc{RydKernel}'s performance to classical alternatives, namely recalling the standard linear and RBF kernels (see Sec.~\ref{supp-subseq:classkernel}).

\subsection{How the kernel is used in a support vector machine \label{supp-subseq:svm}} 
In kernel-based machine learning, a kernel function—classical or quantum—is computed and used within a support vector machine (SVM) for classification, regression and even clustering~\cite{SVM-Hearst1998,Regression-Drucker1996, Clustering-Dhillon2004}. SVMs are a type of supervised learning method that aim to find the best possible boundary, or hyperplane, that separates data points from two different classes. They do this by maximizing the distance, called the margin, between the hyperplane and the closest points from each class—these points are known as support vectors~\cite{Holmes-expressibility2-2022}. SVMs are inherently linear classifiers, but the \textit{kernel trick}—such as using a quantum kernel—allows them to handle nonlinear data by implicitly mapping it to a higher-dimensional space where linear separation is possible. 
SVMs do not natively support multi-class classification and therefore for the purposes of this work, we use binary classification as a benchmark. However, our QKM and SVM generalizes to multi-class classification via standard techniques.

Let us recall that we are working with a classical training data set $\mathcal{S}$ formed of $N_s$ data vectors $\boldsymbol{x}_i \in \mathcal{X}$ along with their true label $y_i \in \mathcal{Y}$. Without loss of generality, we choose $\mathcal{X} = [0,1]^M$ and each vector $\boldsymbol{x}_i = (x_{1,i}, x_{2,i}, \dots, x_{M,i})$ consists of $M$ features.

Given the pairwise inner products between all training samples $\boldsymbol{x}_i \in \mathcal{X}$, the SVM solves the following optimization problem
\begin{equation}
\min_{\boldsymbol{\alpha}} \frac{1}{2} \sum_{i=1}^{N_s} \sum_{j=1}^{N_s} \alpha_i \alpha_j y_i y_j \kappa(\boldsymbol{x}_i, \boldsymbol{x}_j) - \sum_{i=1}^{N_s} \alpha_i, 
\end{equation}
where $y_i \in \mathcal{Y}$ are the corresponding class labels and $\alpha_i$ are Lagrange multipliers. The decision function of an SVM is
\begin{equation}
f(\boldsymbol{x}) = \text{sign} \left( \sum_{i=1}^{N_s} \alpha_i y_i \kappa(\boldsymbol{x}_i, \boldsymbol{x}) + b \right) \,.
\end{equation}
Here, $b$ is the bias term used to determine \textit{the support vectors}. Training an SVM on a dataset of size $N_s$ requires computing $N_s^2$ pairwise inner products to construct the kernel matrix $\kappa$, resulting in at least quadratic runtime. Solving the associated convex quadratic optimization problem may scale as $\mathcal{O}(N_s^2)$ or even $\mathcal{O}(N_s^3)$, depending on the algorithm used. As a result, SVMs are typically suited for small- to moderate-scale problems. However, their convex formulation ensures convergence to a global minimum, offering one of the key advantages of kernel methods. To implement this, the quantum kernel can be integrated into an SVM by using the ``precomputed" kernel option available in Scikit-learn, a python based open-source package for machine learning tasks~\cite{scikit-learn}. 

\subsection{Classical kernel methods used in the paper}
\label{supp-subseq:classkernel}
For our classical benchmarks, we use the linear and radial basis function (RBF) kernels provided in Scikit-learn~\cite{scikit-learn}.  Both kernels are defined below for data points $\boldsymbol{x}_i, \boldsymbol{x}_j \in \mathbb{R}^M$ and specifically for our case $ \boldsymbol{x}_i, \boldsymbol{x}_j\in [0,1]^M$. 
The linear kernel between two input vectors is
\begin{equation}
\kappa_{\text{linear}}(\boldsymbol{x}_i, \boldsymbol{x}_j) = \boldsymbol{x}_i^\top \boldsymbol{x}_j = \sum_{m=1}^M x_{m,i} x_{m,j}.
\end{equation}
The RBF kernel is one of the most commonly used kernels in support vector machines (SVMs), owing to its close relationship with the Gaussian distribution. It is defined as
\begin{align}
\kappa_{\text{RBF}}(\boldsymbol{x}_i, \boldsymbol{x}_j) &= \exp\left(-\gamma \|\boldsymbol{x}_i - \boldsymbol{x}_j\|^2\right) \\ \nonumber &= \exp\left(-\gamma \sum_{m=1}^{M} (x_{m,i} - x_{m,j})^2\right),
\end{align}
where $\gamma > 0$ is a hyperparameter controlling the kernel width. In our case $\gamma = 0.01$. The term $\lVert \boldsymbol{x}_{i} - \boldsymbol{x'}_{j} \rVert^2$ represents the squared Euclidean distance between two feature vectors $\boldsymbol{x}_i$ and $\boldsymbol{x}_j$. It can also be defined in terms of a free parameter $\sigma$, or equivalently, using the parameter $\gamma = \frac{1}{2\sigma^2}$. We observe that the value of the RBF kernel decreases as the distance between $\boldsymbol{x}_i$ and $\boldsymbol{x}_j$ increases, ranging from 1 when $\boldsymbol{x}_i = \boldsymbol{x}_j$ to 0 as the distance approaches infinity. As reported in the literature, support vector classifiers with classical kernels can achieve 100\% accuracy on the IRIS dataset~\cite{iris_53}—a result that is also reflected in our comparison given in the main text.


\end{document}